\documentclass[twocolumn,prb, amsmath,amssymb,letterpaper,longbibliography]{revtex4}

\def\bG{\mathbf{\Gamma}}

\def\bK{\mathbf{K}}
\def\bKp{\mathbf{K'}}
\def\bq{\mathbf{q}}
\def\bk{\mathbf{k}}
\def\bkappa{\bm{\kappa}}
\def\kdotp{\mathbf{k}\cdot\mathbf{p}}

\usepackage{amsmath}
\usepackage{amssymb}
\usepackage{amsfonts}
\usepackage{graphicx}
\usepackage{dcolumn}
\usepackage{float}
\usepackage{bm}
\usepackage{mathrsfs}
\usepackage{txfonts}
\usepackage{tabularx}
\usepackage{bigstrut}
\usepackage[usenames,dvipsnames]{color}
\usepackage{xcolor}
\usepackage{verbatim} 
\usepackage{xfrac}
\usepackage{ulem}

\newcommand{\ForestGreen}      [1]   {\textcolor{ForestGreen}}

\begin{document}

\title{An Effective Model for the Electronic and Optical Properties of Stanene}

\author{Cuauht\'emoc Salazar, Rodrigo A. Muniz, and J. E. Sipe }

\affiliation{Department of Physics, University of Toronto, Toronto ON, M5S 1A7,
Canada }

\date{\today }
\begin{abstract}
  The existence of several 2D materials with heavy atoms has recently
  been demonstrated. The electronic and optical properties of these
  materials can be accurately computed with numerically intensive
  density functional theory methods. However, it is desirable to have
  simple effective models that can accurately describe these
  properties at low energies. Here we present an effective model for
  stanene that is reliable for electronic and optical properties for
  photon energies up to $1.1$ $e$V. For this material, we find that a
  quadratic model with respect to the lattice momentum is the best
  suited for calculations based on the bandstructure, even with
  respect to band warping. We also find that splitting the two
  spin-$\hat{{\bf z}}$ subsectors is a good approximation, which
  indicates that the lattice buckling can be neglected in calculations
  based on the bandstructure.  We illustrate the applicability of the
  model by computing the linear optical injection rates of carrier and
  spin densities in stanene.  Our calculations indicate that an
  incident circularly polarized optical field only excites electrons
  with spin that matches its helicity.
\end{abstract}
\maketitle

\section{Introduction}
\label{sec:intro}

The experimental isolation of single layers of graphene nearly a
decade ago has inspired a search for new 2D
materials\cite{Miro2DAtlas,Gupta_2015}.  Among those that have been
studied are silicene, germanene and stanene
\cite{EzawaReview,Ezawa2015_silicene,Zhao_2016_siliceneR}, zinc-oxide
\cite{Sahoo2016_ZnO_2D}, and the transition metal
dichalcogenides\cite{Wang2012_TMDS,Xu2014_TMDS}.  There is also
substantial research on other elemental 2D materials, including the
remaining elemental crystallogens
\cite{Houssa2014GrapheneLike,Bianco2013,Acun2015_Germanene}, elemental
pnictogens, such as nitrogene \cite{Ozcelik_2015}, phosphorene
\cite{Pereira_2015}, arsenene \cite{Kamal2015_arsenene}, antimonene
\cite{Akturk_2015,Pizzi_2016}, and bismuthene \cite{Aktruk_2016}, as
well as members from other families
\cite{Mannix_2015,2DinjectionCurrent}.  One of the most interesting
materials in this group is stanene, a monolayer of Sn atoms arranged
in a buckled honeycomb lattice. Due to the heavy Sn atoms, the
spin-orbit coupling (SOC) is expected to be strong and to lead to
nontrivial topological properties of the bands that make stanene a 2D
topological insulator \cite{XuPRL2013}. The strong SOC is predicted to
open band gaps of $88$~m$e$V at the $\mathbf{K}$ and
$\mathbf{K^{\prime}}$ points of the Brillouin zone
\cite{XuPRL2013,vandenBroek2014}, and thus the quantum spin Hall
effect, with its characteristic spin polarized edge modes free of
backscattering from non-magnetic impurities, could in principle be
observed at room temperature.  Recently, monolayers of stanene have
been epitaxially grown\cite{FengFeng-snn-exp}, and phase-change laser
ablation techniques\cite{Saxena2016Nature} have been used to produce
few-layer stanene.  Experiments probing high photon energy absorption
properties of few-layered stanene have also been reported
\cite{Chaudhary2016}.

While the electronic and optical properties of crystalline materials
can be studied with modern \textit{ab initio} methods, the numerical
task can be challenging. It is thus desirable to have simple effective
models that reliably reproduce the basic properties of materials, at
least over energy ranges of interest. In order to compute electronic
and optical properties from an effective model, it is necessary to
know the Hamiltonian and the Lax connection\footnote{Notice that in
  this paper we refer to the connections in the Brillouin zone
  introduced by Melvin Lax\cite{LaxBook}. They should not be confused
  with the connections related to Lax pairs introduced by Peter Lax.},
which gives important geometric information about the basis of the
quantum states\cite{kpBook} in the model. Two of the most common types
of effective models for crystals are tight-binding and
${\bf k}\cdot{\bf p}$ models.

In tight-binding models, the basis of states is defined in terms of
a set of Wannier functions that are exponentially localized in space;
it is always possible to obtain such a set of functions for a block
of electronic bands with vanishing total Chern number that do not
cross others \cite{marzari97,marzari12}. The Hamiltonian and the
Lax connection are respectively expressed in terms of hopping parameters
and dipole matrix elements. The hopping parameters can be inferred
from bandstructure properties, obtained either from experiments or
from first-principle calculations. In contrast, the Lax connection 
parameters are harder to deduce since they are usually obtained from
electronic and optical properties. When the Wannier functions are
well localized, the overlap between them -- and consequently the matrix
elements for any operator -- can be restricted to only nearest neighbor
atomic sites; the model is then usually simple and has relatively
few parameters that need to be inferred. However, if the Wannier functions
at sites further apart have a considerable overlap, the number of
free parameters increases significantly. While this is not a major
problem for determining hopping parameters, it leads to a large number
of dipole parameters that are hard to fit.

In $\kdotp$ models, the basis of states consists of the periodic parts
$u_{\ell{\bf q}}\left({\bf r}\right)$ of Bloch wavefunctions
$\psi_{\ell\mathbf{q}}\left(\mathbf{r}\right)=e^{i\mathbf{q}\cdot\mathbf{r}}u_{\ell\mathbf{q}}\left(\mathbf{r}\right)/\sqrt{\left(2\pi\right)^{D}}$
for a set of bands $\ell$ at a reference point $\bq$ in the Brillouin
zone (BZ) of dimension $D$. Since the basis is independent of the
lattice momentum $\bk$, the Lax connection is null for a
${\bf k}\cdot{\bf p}$ model, which simplifies the calculation of
electronic and optical properties. However, ${\bf k}\cdot{\bf p}$
models also have drawbacks. For instance, the Hamiltonian has a fixed
form that is quadratic in the lattice momentum $\bk$, but its free
parameters are only associated with the linear terms in $\bk$, as the
quadratic term is related to the electron bare mass. Because of that,
the only way to introduce more parameters in the Hamiltonian is to
increase the number of bands in the model, even if the additional
bands are irrelevant except for aiding in the fitting of the band
energies of interest. Also, since the periodic functions depend on
$\bk$, the basis needs to include the states of several bands at the
reference $\bq$ point in order to span the state of a single band at
other $\bk$ points in the BZ. Thus ${\bf k}\cdot{\bf p}$ models for
the whole Brillouin zone usually include several bands, but describe
only a few of them accurately, a fact that increases the number of
parameters to be inferred. Moreover, the accuracy of the
states\cite{CardonaPollak,Richard04,kpBook} at a point $\bk$ in the BZ
decreases with the distance from the reference point $\bq$, and since
results are usually reported without a standard measure of the error,
it is not possible to know exactly where the approximation becomes
unacceptable.

In this article, we develop an effective model for stanene that is
similar to a $\kdotp$ model but that is free of the drawbacks pointed
out in the previous paragraph. We keep track of the accuracy of the
eigenstates, and the free parameters of the Hamiltonian are not
restricted to the linear terms in the lattice momentum $\bk$. Starting
from an \textit{ab initio} set of wavefunctions, we expand the
eigenstates at a region of the BZ in terms of the states at a
reference point $\bq$ in that region. For a finite set of bands, this
expansion is not unitary, as the basis set is incomplete. In order to
preserve unitarity, we approximate this expansion by a unitary
transformation\cite{JLC13} using a singular value decomposition (SVD),
the singular values of which provide a measure of the accuracy of the
eigenstates. This transformation allows the same basis to be used for
a region of the BZ, so the Lax connection is null as desired. A Taylor
expansion of the Hamiltonian matrix written in this basis with respect
to the lattice momentum $\bk$ then gives the free parameters of our
model. For stanene we use three regions in the BZ, around the points
$\mathbf{K}$, $\mathbf{K^{\prime}}$, and $\boldsymbol{\Gamma}$. We
obtain an effective model that is accurate for transition energies up
to $1.1$~eV, with a quadratic expansion for each reference point. We
find that the band warping is well accounted for by a quadratic model,
and that a cubic model does not improve upon it significantly. We also
find that neglecting some small parameters leads to the separation of
the spin sectors in our model; such approximation is accurate within a
tolerance corresponding to the room temperature energy.

To illustrate the applicability of our model, we compute the
one-photon injection rate coefficients for carrier and spin densities
in stanene. We predict that an incident circularly polarized optical
field with photon energy close to the gap only excites electrons with
spins that match the helicity of the optical field. This result
suggests the possibility of employing stanene in optically-controlled
spin pump applications.

The outline of this article is as follows: In Sec. \ref{sec:model}
we present the procedure to obtain the effective model; in Sec. \ref{sec:effmodel}
we apply it to stanene and analyze the accuracy of the eigenstates
and the eigenenergies, including the band warping. In Sec. \ref{sec:conducti}
we use our model to compute linear optical absorption rates of stanene.
We end with a discussion of our results in Sec. \ref{sec:conclusion}.

\section{Method for deriving effective models }
\label{sec:model}

Bloch's theorem asserts that the eigenstates $\psi_{\ell\mathbf{k}}\left(\mathbf{r}\right)$
of a periodic Hamiltonian function ${\cal H}\left(\mathbf{r},-i\hbar\boldsymbol{\nabla}\right)={\cal H}\left(\mathbf{r}+\mathbf{R},-i\hbar\boldsymbol{\nabla}\right)$,
where $\mathbf{R}$ is a lattice vector, can be written as 
\begin{align}
\psi_{\ell\mathbf{k}}\left(\mathbf{r}\right)=\frac{1}{\sqrt{\left(2\pi\right)^{D}}}e^{i\mathbf{k}\cdot\mathbf{r}}u_{\ell\mathbf{k}}\left(\mathbf{r}\right),
\end{align}
where $u_{\ell\mathbf{k}}\left(\mathbf{r}\right)=u_{\ell\mathbf{k}}\left(\mathbf{r}+\mathbf{R}\right)$
are periodic functions. In typical \textit{ab initio} calculations,
a very large number of basis functions $\tilde{u}_{a\mathbf{k}}\left(\mathbf{r}\right)$,
which usually consist of plane waves or atomic orbitals, are used to
specify  the Bloch Hamiltonian ${\cal H}\left(\mathbf{r},-i\hbar\boldsymbol{\nabla}+\hbar\mathbf{k}\right)$
by  the matrix elements
\begin{align}
\tilde{H}_{ab\mathbf{k}} &=\langle \tilde{u}_{a\bk} | {\cal H}_{\bk}  | \tilde{u}_{b\bk}\rangle \notag \\
&\equiv \Omega_{\mathrm{uc}}^{-1}\int_{\mathrm{uc}}d\mathbf{r}\,\tilde{u}_{a\mathbf{k}}^{\ast}\left(\mathbf{r}\right){\cal H}\left(\mathbf{r},-i\hbar\boldsymbol{\nabla}+\hbar\mathbf{k}\right)\tilde{u}_{b\mathbf{k}}\left(\mathbf{r}\right),
\end{align}
where $\Omega_{\mathrm{uc}}$ is the volume of the unit cell. The
Hamiltonian matrix $\tilde{H}_{\mathbf{k}}$ consisting of these
elements is then diagonalized, and provides the eigenstates and
eigenenergies corresponding to each electronic band $\ell$ at the
lattice momentum $\mathbf{k}$. We denote the diagonalized matrix by
$H_{\bk}$. If the large set of basis functions in the \textit{ab
  initio} calculation are taken to be the same for different lattice
momenta, say ${\bf q}$ and $\bk$, we can compute the overlap matrix
between states, ${\cal W}_{\bk;\bq}$,  with matrix elements
\begin{equation}
{\cal W}_{m\ell\bk;\bq}=\langle u_{m\mathbf{q}}|u_{\ell\mathbf{k}}\rangle=\Omega_{\mathrm{uc}}^{-1}\int_{\mathrm{uc}}d\mathbf{r}\ u_{m\mathbf{q}}^{\ast}\left(\mathbf{r}\right)u_{\ell\mathbf{k}}\left(\mathbf{r}\right).\label{eq:overlap}
\end{equation}
The overlap matrix allows us to decompose the states
$u_{\ell{\bf k}}\left({\bf r}\right)$ at $\bk$ in terms of those at
the reference point ${\bf q}$ in the BZ and to use the states
$\{u_{m{\bf q}}\left({\bf r}\right)\}$ as a basis for any $\bk$ point
in the region of the BZ around $\mathbf{q}$.  In order to have a
simple effective model, it is desirable to include only a small number
of bands in the basis set. However, if only a few functions
$u_{m{\bf q}}\left({\bf r}\right)=\langle {\bf r}\,|\,u_{m{\bf q}}\rangle$
are included in the basis, even the states
$u_{m{\bf k}}\left({\bf r}\right)=\langle {\bf r}\,|\,u_{m{\bf k}}\rangle$ corresponding to the same block of
bands at other $\bk$ point in the BZ neighborhood might not be
completely spanned by them. This means that the overlap matrix
${\cal W}_{\bk;\bq}$ might not be unitary when restricted to a small
set of bands. Here we ensure the unitarity of the model by replacing
${\cal W}_{\bk;\bq}$ with a unitary matrix based on its singular value
decomposition (SVD). In the remaining of this discussion we drop the
subindex indicating the reference $\bq$ point in the BZ where it does
not lead to confusion. In its singular form, the overlap matrix
${\cal W}_{\bk}$ is written as
\begin{align}
{\cal W}_{\bk}=U_{\mathbf{k}}\Sigma_{\mathbf{k}}V_{\mathbf{k}}^{\dagger},\label{eq:SVDoverlap}
\end{align}
where $U_{\mathbf{k}}$ and $V_{\mathbf{k}}$ are unitary matrices, and
$\Sigma_{\mathbf{k}}$ is a diagonal matrix with its elements as the
singular values.  If ${\cal W}_{\mathbf{k}}$ were a unitary matrix,
$\Sigma_{\mathbf{k}}$ would be the identity matrix $I$, thus a simple
``unitary approximation'' to ${\cal W}_{\mathbf{k}}$ is to replace
$\Sigma_{\mathbf{k}}$ with the identity matrix as
\begin{align}
{\cal W}_{\mathbf{k}}\to W_{\mathbf{k}}\equiv U_{\mathbf{k}}V_{\mathbf{k}}^{\dagger}.\label{eq:uapprox}
\end{align}
An obvious measure for the accuracy of this approximation is the difference
$\mathbf{I}-\Sigma_{\mathbf{k}}$. For each $\bk$ in a region around the reference
$\bq$ point in the BZ, the approximate unitary overlap matrix $W_{\mathbf{k}}$
allows the expansion of the states $|u_{\ell\mathbf{k}}\rangle$ in terms
of the basis $|u_{m\mathbf{q}}\rangle$ as
\begin{equation}
\left|u_{\ell\mathbf{k}}\right\rangle =\sum_{m}W_{m\ell\mathbf{k}}\left|u_{m\mathbf{q}}\right\rangle .
\label{eq:uWu}
\end{equation}
The next step is to use the above equation to write the Hamiltonian
matrix $H_{\bk}$ for each $\bk$ in terms of the states
$\left|u_{m\mathbf{q}}\right\rangle$ at the reference $\bq$ point in
the BZ.  Note that the $\left|u_{\ell\mathbf{k}}\right\rangle $ are
the eigenstates of the Hamiltonian matrix $H_{\mathbf{k}}$, at lattice
momentum $\mathbf{k}$ is 
\begin{equation}
\begin{array}{rl}
H_{m\ell\mathbf{k}} & =\left\langle u_{m\mathbf{k}}\right|{\cal H}_{\mathbf{k}}\left|u_{\ell\mathbf{k}}\right\rangle =\delta_{m\ell}E_{\ell\mathbf{k}}.\end{array}
\end{equation}
We write the elements of the Hamiltonian matrix for lattice momentum
$\bk$ expressed in the $\left|u_{\ell\mathbf{q}}\right\rangle $ basis
as
\begin{equation}
\begin{array}{r}
\bar{H}_{m\ell\mathbf{k}}=\left\langle u_{m\mathbf{q}}\right|{\cal H}_{\mathbf{k}}\left|u_{\ell\mathbf{q}}\right\rangle \end{array},
\end{equation}
and using Eq.~\eqref{eq:uWu},
the matrix $\bar{H}_{\mathbf{k}}$ is  related to $H_{\mathbf{k}}$ through the unitary matrix $W_{\mathbf{k}}$ that performs the change of basis 
\begin{equation}
\bar{H}_{\mathbf{k}}=W_{\mathbf{k}}E_{\mathbf{k}}W_{\mathbf{k}}^{\dagger},
\label{eq:why}
\end{equation}
where $E_{\mathbf{k}}$ is a diagonal matrix with diagonal elements
$E_{\ell\mathbf{k}}$.
Since the basis of states
$\left\{\left|u_{m\mathbf{q}}\right\rangle\right\}$ is independent of
$\mathbf{k}$, its Lax connection vanishes,
$\bar{ \boldsymbol{\xi}}_{m\ell\mathbf{k}}\equiv i\left\langle
  u_{m\mathbf{q}}\right|\boldsymbol{\nabla}_{\mathbf{k}}\left|u_{\ell\mathbf{q}}\right\rangle
=\mathbf{0}$.  Consequently, such a basis is suitable for expanding
the Hamiltonian matrix $\bar{H}_{\mathbf{k}}$ around $\mathbf{q}$
simply as
\begin{equation}
\bar{H}_{\mathbf{k}}=\bar{H}_{\mathbf{q}}+\bkappa\cdot\left.\boldsymbol{\nabla}_{\mathbf{k}}\bar{H}_{\mathbf{k}}\right|_{\mathbf{k}=\mathbf{q}}+{\cal O}\left(\kappa^{2}\right)+{\cal O}\left(\kappa^{3}\right)\dots,\label{eq:effH}
\end{equation}
where $\bkappa=\mathbf{k}-\mathbf{q}$. If the basis were dependent
on the lattice momentum $\bk$, the expansion would include a correction
given by the Lax connection. 

In summary, the overlap matrix ${\cal W}_{\mathbf{k}}$ from an
\textit{ab initio} calculation is replaced by its unitary
approximation $W_{\mathbf{k}}$, the diagonalized Hamiltonian is
written in a basis that is independent of the lattice momentum $\bk$,
and a Taylor expansion of its matrix elements gives the free
parameters in our model. We now turn to the application of this
procedure to stanene.

\section{Effective model for stanene }

\begin{figure}[htb!]
\includegraphics[width=0.95\columnwidth]{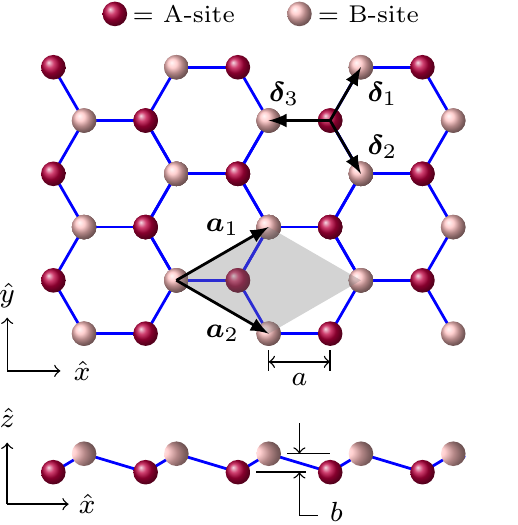}
\caption[]{ (Color online) Hexagonal lattice of stanene with Sn atoms at A and
B sites. The lattice vectors are denoted by $\bm{a}_{1}$ and $\bm{a}_{2}$,
and we show the unit cell with a gray rhombus. The interatomic distance projected
on the plane is $a$, and along the vertical direction it is $b$,
so the interatomic distance is $\sqrt{a^{2}+b^{2}}$.}
\label{fig:lattices} 
\end{figure}

We start by obtaining the electronic wavefunctions from a
first-principles calculation, in the framework of Density Functional
Theory (DFT) and the Local Density Approximation (LDA), using the
freely available ABINIT code\cite{Gonze20092582,Gonze2016106}. The
wavefunctions are expanded in a basis of planewaves; the size of the
basis is determined by a kinetic-energy cutoff of $653$~$e$V,
corresponding to 6166 planewaves. The crystal (ionic) potential is
modeled using the Optimized Norm-Conserving Vanderbilt
Pseudopotentials (ONCVP) \cite{HamannPSP8}; we take 14 out of the 50
Sn electrons as valence electrons, and the others are assumed
clamped. We converge the ground-state total energy up to $2.7$~m$e$V,
leading to a $12\times12$ ${\bf k}$-point mesh. Since we simulate the
Sn monolayer with a supercell model, we introduce an interlayer vacuum
space of $11.42$~\AA , such that spurious inter-layer interactions are
negligible; with this amount of vacuum space, the total energy remains
unchanged within $2.7$~m$e$V if the vacuum space is
incremented. Relaxing the atomic positions leaves the atoms at the
$\left(x,y\right)$ coordinates of a honeycomb lattice, i.e., one Sn
atom at $\left(0,0\right)$ and another at
$\left(\bm{a}_{1}+\bm{a}_{2}\right)/3$.  The lattice vectors are
$\bm{a}_{1}=a\left(3\hat{\bm{x}}+\sqrt{3}\hat{\bm{y}}\right)/2$ and
$\bm{a}_{2}=a\left(3\hat{\bm{x}}-\sqrt{3}\hat{\bm{y}}\right)/2$, where
$a=2.66$~\AA{} is the interatomic distance projected on the plane. The
relaxation of the $z$-coordinates leads to out-of-plane coordinates
$\pm0.418$~\AA , giving rise to a ``buckling distance'' of
$b=0.836$~\AA , such that the interatomic distance is
$\sqrt{a^{2}+b^{2}}$. This low buckling has been shown to enhance the
overlap between $\pi$ and $\sigma$ orbitals, leading to an equilibrium
configuration in materials where the $\pi$-$\pi$ bonding is relatively
weak\cite{Cahangirov2009,XuPRL2013}. The first nearest neighbour
vectors are
$\bm{\delta}_{1}= \tfrac{a}{2}
(\hat{\bm{x}}+\sqrt{3}\hat{\bm{y}})+b\hat{{\bf z}}$,
$\bm{\delta}_{2}=\tfrac{a}{2}
(\hat{\bm{x}}-\sqrt{3}\hat{\bm{y}})+b\hat{{\bf z}}$ and
$\bm{\delta}_{3}=-a\hat{\bm{x}}+b\hat{{\bf z}}$.  In Fig.
\ref{fig:lattices}, we show the crystal lattice of stanene.

\begin{figure}[htb!]
\includegraphics[width=0.95\columnwidth]{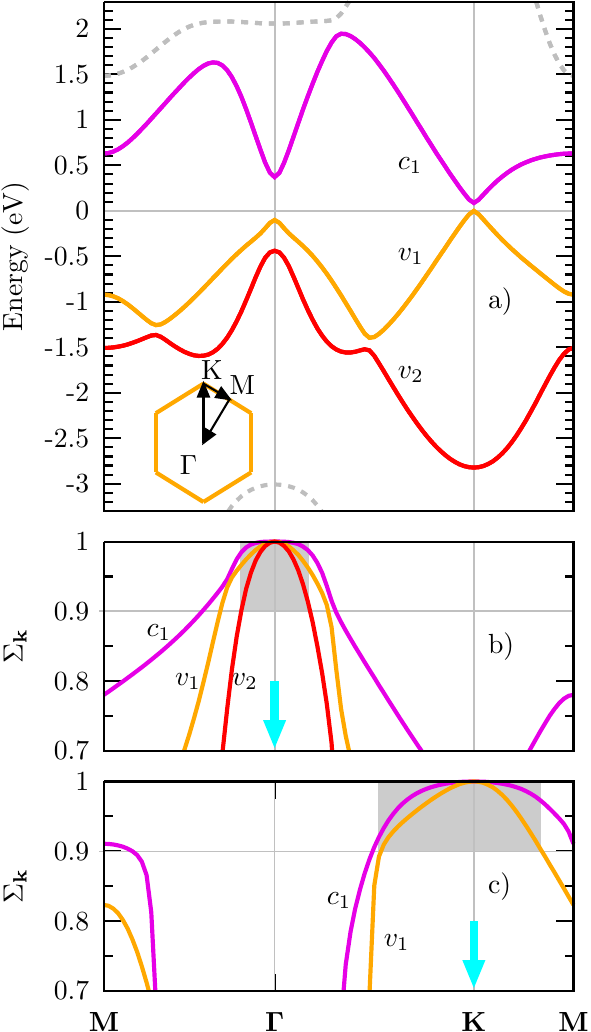}
\caption[]{ (Color online) a) The \textit{ab initio} bandstructure of
  stanene with the bands included in our effective model
  highlighted. The band gaps at $\mathbf{K}$ and $\mathbf{K}^{\prime}$
  have a value of 88~meV.  At $\Gamma$, the minimal transition is at
  0.472~eV and the second is at 0.808~eV. All bands are doubly (spin)
  degenerate. The dashed (gray) bands are not described by our
  model. b) The singular values (the elements of the diagonal matrix
  $\Sigma_{\mathbf{k}}$, Eq.~\eqref{eq:SVDoverlap}) with $\bq=\bG$ as
  the reference point. The shaded area indicates the region where 
  the singular values are all greater than $0.9$, and the unitary
  approximation $\Sigma_{\mathbf{k}}\to\bm{I}$ is acceptable. c) Same
  as b), but for $\mathbf{q}=\mathbf{K}$ as the reference point.}
\label{fig:bandssingval} 
\end{figure}

\begin{figure}[htb!]
\includegraphics[width=0.95\columnwidth]{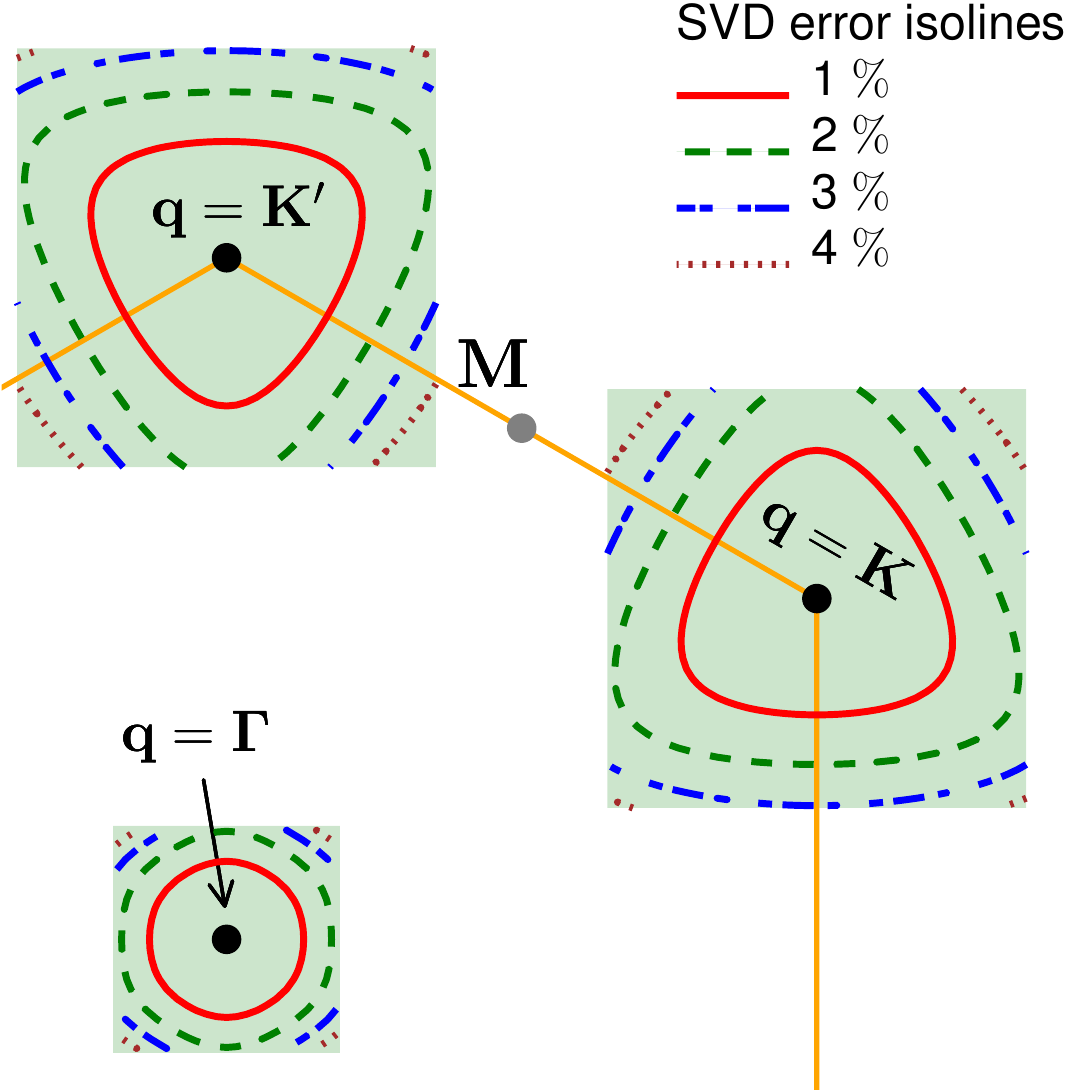}
\caption[]{ (Color online) Figure of merit of the unitary
  approximation $\Sigma_{\mathbf{k}}\to\bm{I}$, as defined by
  Eq.~\eqref{eq:figmerit}, for the three regions of the BZ centered at
  the reference points $\bq=\{\bm{\Gamma},\mathbf{K},\mathbf{K}'\}$;
  each $\bq$ is marked with black dots. The yellow lines connecting
  all contiguous pairs of $\bK$ and $\bKp$ points indicate boundaries
  of the first Brillouin zone (cf. inset of
  Fig.~\ref{fig:bandssingval}, a)).}
\label{fig:SVDerror} 
\end{figure}

With these structural parameters we proceed to obtain the bandstructure
along the typical $\mathbf{M}\bm{\Gamma}\mathbf{K}\mathbf{M}$ path
(Fig.~\ref{fig:bandssingval}) in the BZ. All the bands are spin
degenerate since stanene has both time-reversal and inversion symmetries.
As shown in Fig.~\ref{fig:bandssingval}, the bandstructure of stanene
has gaped Dirac cones at the $\mathbf{K}$ and $\mathbf{K}^{\prime}$
points with a gap of $0.088\ e$V. At $\bm{\Gamma}$, the first transition
occurs at $0.472\ e$V~ and the next one at $0.808\ e$V. At $\mathbf{M}$
the first transition is at $1.55\ e$V; hence we ignore that region
of the BZ in our model, as we are focusing on energies up to $1.1\ e$V
in this paper. Our effective model contains only states with lattice
momentum around the $\mathbf{K}$, $\mathbf{K}^{\prime}$, and $\bm{\Gamma}$
points; it includes 6 bands around the $\bm{\Gamma}$ point and only
4 bands around $\mathbf{K}$ and $\mathbf{K}^{\prime}$ points.

\subsection{Accuracy of the approximation for the states}
\label{sec:effmodel}

Once the \textit{ab initio} wavefunctions are computed, we proceed to
obtain the overlap matrix (Eq.~\eqref{eq:overlap}) between the
periodic functions at the reference point $\bq$ and the other points
in its neighborhood in the BZ; this is done for each of the regions of
interest in the BZ, namely the regions around the reference points
$\mathbf{K}$, $\mathbf{K}^{\prime}$, and $\bm{\Gamma}$. The overlap
matrices of the \textit{ab initio} wavefunctions can be approximated
by unitary matrices based on singular value decompositions according
to Eq.~\eqref{eq:uapprox}. In order to determine the region of the BZ
where this approximation is accurate, in Fig.~\ref{fig:bandssingval}
we plot the elements of the diagonal matrix $\Sigma_{\mathbf{k}}$ (the
singular values) for the reference points $\mathbf{K}$ and
$\bm{\Gamma}$; the results for the $\mathbf{K}^{\prime}$ point are
similar to those of $\mathbf{K}$. In Fig.~\ref{fig:bandssingval}, we
also highlight the regions where each element of $\Sigma_{\mathbf{k}}$
is greater than 0.9, which is taken as our tolerance for the
approximation in Eq.~\eqref{eq:uapprox}. Notice that the highlighted
regions encompass every point on the BZ where optical transitions with
photon energies below $1.1\ e$V are possible.

In order to have a measure of the accuracy of the states that is easier
to be visualized, we define a figure of merit 
\begin{align}
\delta_{\Sigma}\left(\mathbf{k}\right)=n^{-1}\,\sqrt{\mathrm{Tr}\left(\Sigma_{\mathbf{k}}-\bm{I}\right)^{2}},\label{eq:figmerit}
\end{align}
where $n$ is the number of bands included in the model. In Fig.~\ref{fig:SVDerror}
we present the figure of merit $\delta_{\Sigma}\left(\mathbf{k}\right)$
for the three regions of interest in the BZ. We notice that the error
indicated by $\delta_{\Sigma}\left(\mathbf{k}\right)$ is lower than
5\% for large neighborhoods around the reference points.

\subsection{Hamiltonian matrices }
\label{sec:effmodelMat}

Having established the regions where the approximation of the states
is valid, we now turn to the approximation of the Hamiltonian matrix.
We expand the matrix elements of the Hamiltonian $\bar{H}_{{\bf k}}$
directly as in Eq.~\eqref{eq:effH}, and report the results below.
Since we use a basis independent of the lattice momentum for the
neighborhood of the BZ around each reference point, the Lax connection
is null for each of these neighborhoods,
$\bar{\boldsymbol{\xi}}_{ab\mathbf{k}}=0$.

\subsubsection{$\mathbf{K}$ and $\mathbf{K}^{\prime}$ points}
\label{sec:effmodelK} 

The valleys around the $\bK$ and $\mathbf{K}^{\prime}$ points are
similar in our model, so we present the matrices associated with each
of them together, and use the valley parameter $\tau=1$ to refer to
$\bK$ and $\tau=-1$ to refer to ${\bf K}^{\prime}$. At the $\bK$ and
$\mathbf{K}^{\prime}$ points, the wavefunctions have a predominant
character of $p_{z}$ orbitals located at an atom in the unit cell.  We
use $s_{i}$ and $\sigma_{i}$ to respectively denote the Pauli matrices
in the spin and sublattice sectors; here $i=\left\{ 0,x,y,z\right\}$,
as we adopt the convention of denoting the identity as the zeroth
Pauli matrix. In this notation, the Hamiltonian is written in terms of
the matrices $s_{i}\otimes\sigma_{j}$.

Up to linear order in the lattice momentum $\boldsymbol{\kappa}=\mathbf{k}-\mathbf{q}$,
where $\mathbf{q}=\left\{ \bK,\mathbf{K}^{\prime}\right\} $, we find
\begin{align}
\bar{H}_{\tau\boldsymbol{\kappa}}^{\left(1\right)}= & \Delta_{K}\left(-\tau s_{z}\otimes\sigma_{z}+s_{0}\otimes\sigma_{0}\right)+\zeta_{K}^{\left(1\right)}as_{0}\otimes\left(\kappa_{x}\sigma_{x}+\tau\kappa_{y}\sigma_{y}\right)\notag\label{eq:H1K}\\
 & -\lambda_{K}^{\left(1\right)}a\left(\kappa_{y}s_{x}-\kappa_{x}s_{y}\right)\otimes\sigma_{z},
\end{align}
where in the first term we add an energy shift $\Delta_{K}$ such
that the top of the valence band is at zero energy. The quadratic
terms in $\boldsymbol{\kappa}$ are 
\begin{align}
\bar{H}_{\tau\boldsymbol{\kappa}}^{\left(2\right)}= & -\zeta_{K}^{\left(2\right)}a^{2}s_{0}\otimes\left[\tau\kappa_{x}\kappa_{y}\sigma_{x}+\frac{1}{2}\left(\kappa_{x}^{2}-\kappa_{y}^{2}\right)\sigma_{y}\right]\notag\label{eq:H2K}\\
 & -v_{K}^{\left(2\right)}a^{2}\left|\kappa\right|^{2}s_{0}\otimes\sigma_{0}+\vartheta_{K}^{\left(2\right)}a^{2}\tau\left|\kappa\right|^{2}s_{z}\otimes\sigma_{z}\\
 & +\eta_{K}^{\left(2\right)}a^{2}\tau\left[\left(\kappa_{x}^{2}-\kappa_{y}^{2}\right)s_{x}-2\kappa_{x}\kappa_{y}s_{y}\right]\otimes\sigma_{z},\notag
\end{align}
where the values of the parameters are shown in Table
\ref{tab:paramK}.  Neglecting the relatively small parameters
$\lambda_{K}^{\left(1\right)}$ and $\eta_{K}^{(2)}$ leads to a
separation of the spin subsectors, since without them
$\bar{H}_{\tau{\bf k}}^{\left(1\right)}$ and
$\bar{H}_{\tau{\bf k}}^{\left(2\right)}$ do not have terms with
$s_{x}$ and $s_{y}$, the only matrices with cross-spin elements. The
spin separation is expected for lattices without buckling, and it
indicates that the lattice buckling can be neglected in calculations
involving $\bk$ close to the expansion point $\bq$.

The parameters $v_{K}^{\left(2\right)}$ and
$\vartheta_{K}^{\left(2\right)}$ can also be neglected, and the three
parameters $\Delta_{K}$, $\zeta_{K}^{\left(1\right)}$ and
$\zeta_{K}^{\left(2\right)}$ are the only ones needed for our model to
give band energies that match those from DFT within a tolerance of
room temperature energy. We nevertheless report the negligible
parameters $\lambda_{K}^{\left(1\right)}$, $\eta_{K}^{(2)}$,
$v_{K}^{\left(2\right)}$ and $\vartheta_{K}^{\left(2\right)}$, because
their physical significance can be identified with the help of a
$p_{z}$-orbital tight-binding model, as we discuss in the Appendix.
Finally, we provide an analytical expression for the band
  energies around the $\mathbf{K}$ and $\mathbf{K}^{\prime}$ points
  obtained from our effective model. Neglecting the small parameters
  mentioned in the previous paragraph, we have 
\begin{subequations}
\begin{align}
E_{\tau\boldsymbol{\kappa}}^{\pm} & =\Delta_{K}\pm\sqrt{\Delta_{K}^{2}+\mathcal{X}_{\boldsymbol{\kappa}}^{2}+\mathcal{Y}_{\boldsymbol{\kappa}}^{2}},\\
\mathcal{X}_{\boldsymbol{\kappa}} & =a\kappa_{x}\left(\zeta_{K}^{\left(1\right)}-\tau\zeta_{K}^{\left(2\right)}a\kappa_{y}\right), \\
\mathcal{Y}_{\boldsymbol{\kappa}} & =\zeta_{K}^{\left(1\right)}a\kappa_{y}-\frac{1}{2}\tau\zeta_{K}^{\left(2\right)}a^{2}\left(\kappa_{x}^{2}-\kappa_{y}^{2}\right),
\end{align}
\end{subequations} where the positive and negative signs of the square
root correspond to the conduction and valence bands respectively.

\begin{table}[htb]
\begin{tabular}{|c|c|c|}
\hline 
\multicolumn{3}{|c|}{All values in eV }\tabularnewline
\hline 
 $\Delta_{K}=0.044$  & $\zeta_{K}^{\left(1\right)}=0.67$  & $\zeta_{K}^{\left(2\right)}=0.33$ \tabularnewline
 & $\lambda_{K}^{\left(1\right)}=0.03$  & $v_{K}^{\left(2\right)}=0.03$ \tabularnewline
 &  & $\vartheta_{K}^{\left(2\right)}=0.03$ \tabularnewline
 &  & $\eta_{K}^{\left(2\right)}=0.02$ \tabularnewline
\hline 
\end{tabular}\caption{Parameter values of the models for the $\bK$ and $\bKp$ valleys
in the BZ, see Eqs.~\eqref{eq:H1K} and \eqref{eq:H2K}. The parameters
$\lambda_{K}^{\left(1\right)}$, $\eta_{K}^{(2)}$, $v_{K}^{\left(2\right)}$
and $\vartheta_{K}^{\left(2\right)}$ can be neglected without significant
changes in the band energies. Neglecting the parameters $\lambda_{K}^{\left(1\right)}$
and $\eta_{K}^{(2)}$ alone already leads to a separation of the spin
subsectors. }
\label{tab:paramK} 
\end{table}

\subsubsection{$\boldsymbol{\Gamma}$ point } 
\label{sec:effmodelGam}

At the $\bG$ point, the wavefunctions cannot be easily associated with
a sublattice, but they can still be identified according to spin, so
we continue using $s_{i}$ to denote the Pauli matrices acting on the
spin sector of the Hilbert space. Up to linear order in the lattice
momentum, here $\boldsymbol{\kappa}=\mathbf{k}-\mathbf{q}=\mathbf{k}$
since $\mathbf{q}=\boldsymbol{\Gamma}$, we find
\begin{equation}
\begin{array}{rl}
\label{eq:H1G}\bar{H}_{\Gamma\boldsymbol{\kappa}}^{\left(1\right)}= & s_{0}\otimes\begin{bmatrix}E_{\Gamma}^{\left(c\right)} & 0 & 0\\
0 & E_{\Gamma}^{\left(v1\right)} & 0\\
0 & 0 & E_{\Gamma}^{\left(v2\right)}
\end{bmatrix}+a\kappa_{x}s_{0}\otimes\begin{bmatrix}0 & \zeta_{\Gamma1}^{\left(1\right)} & \zeta_{\Gamma2}^{\left(1\right)}\\
\zeta_{\Gamma1}^{\left(1\right)} & 0 & 0\\
\zeta_{\Gamma2}^{\left(1\right)} & 0 & 0
\end{bmatrix}\\
 & +a\kappa_{y}s_{z}\otimes\begin{bmatrix}0 & -i\zeta_{\Gamma1}^{\left(1\right)} & i\zeta_{\Gamma2}^{\left(1\right)}\\
i\zeta_{\Gamma1}^{\left(1\right)} & 0 & 0\\
-i\zeta_{\Gamma2}^{\left(1\right)} & 0 & 0
\end{bmatrix},
\end{array}
\end{equation}
while the quadratic terms in $\boldsymbol{\kappa}$ are 
\begin{equation}
\begin{array}{rl}
\label{eq:H2G}\bar{H}_{\Gamma\boldsymbol{\kappa}}^{\left(2\right)}= & \frac{1}{2}a^{2}\left|\kappa\right|^{2}s_{0}\otimes\begin{bmatrix}v_{\Gamma c}^{\left(2\right)} & 0 & 0\\
0 & -v_{\Gamma1}^{\left(2\right)} & 0\\
0 & 0 & -v_{\Gamma2}^{\left(2\right)}
\end{bmatrix}\\
 & +\frac{1}{2}a^{2}\left(\kappa_{x}^{2}-\kappa_{y}^{2}\right)s_{0}\otimes\begin{bmatrix}0 & 0 & 0\\
0 & 0 & \zeta_{\Gamma v}^{\left(2\right)}\\
0 & \zeta_{\Gamma v}^{\left(2\right)} & 0
\end{bmatrix}\\
 & +a^{2}\kappa_{x}\kappa_{y}s_{z}\otimes\begin{bmatrix}0 & 0 & 0\\
0 & 0 & i\zeta_{\Gamma v}^{\left(2\right)}\\
0 & -i\zeta_{\Gamma v}^{\left(2\right)} & 0
\end{bmatrix}.
\end{array}
\end{equation}
The values of the parameters are presented in Table~\ref{table:paramG}.
Here we have omitted negligible parameters. The parameters reported
constitute the minimum set necessary to describe the energies with
an accuracy equivalent to room temperature when compared to the bands
from DFT. Notice that the model for the valley at the $\bG$ point
can also be separated in two spin sectors. 

\begin{table}[htb!]
\begin{tabular}{|c|c|c|}
\hline 
\multicolumn{3}{|c|}{All values in $e$V }\tabularnewline
\hline 
$E_{\Gamma}^{c}=0.37$  & $\zeta_{\Gamma1}^{\left(1\right)}=1.23$  & $v_{\Gamma c}^{\left(2\right)}=0.34$ \tabularnewline
$E_{\Gamma}^{v1}=-0.10$  & $\zeta_{\Gamma2}^{\left(1\right)}=1.16$  & $v_{\Gamma1}^{\left(2\right)}=0.45$ \tabularnewline
$E_{\Gamma}^{v2}=-0.44$  &  & $v_{\Gamma2}^{\left(2\right)}=0.34$ \tabularnewline
 &  & $\zeta_{\Gamma v}^{\left(2\right)}=0.35$ \tabularnewline
\hline 
\end{tabular}\caption{Parameter values of the model for the $\bG$ valley, see Eqs.~\eqref{eq:H1G}
and \eqref{eq:H2G}. Negligible parameters are omitted. \label{table:paramG} }
\end{table}

\subsection{Accuracy of the energies }
\label{sec:bands}

\begin{figure}[htb!]
\includegraphics[width=0.95\columnwidth]{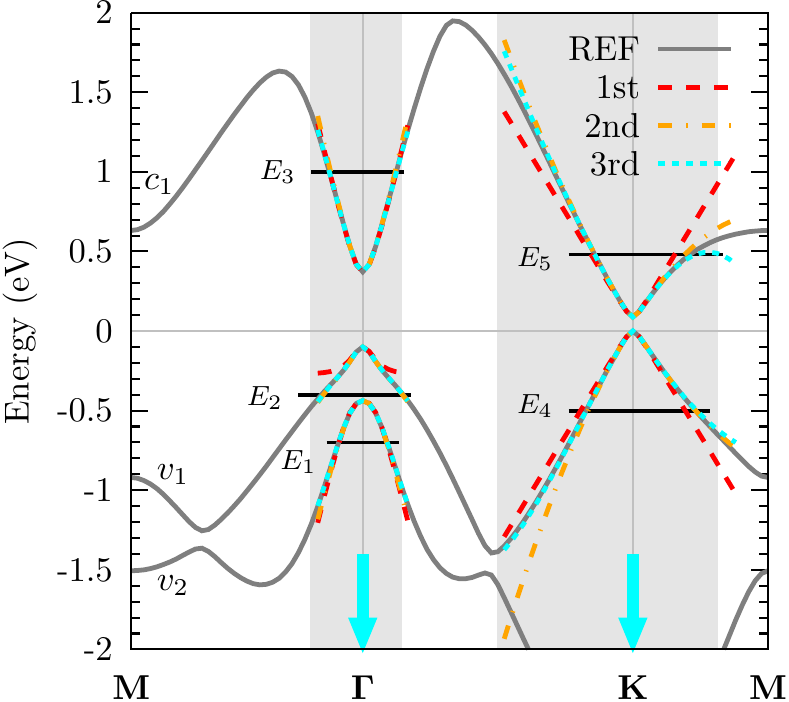}
\caption[]{ (Color online) Comparison of the band energies obtained from the
Taylor expansions of the Hamiltonian matrices (dashed lines) in Sec.~\ref{sec:effmodelMat}
with those from the \textit{ab initio} calculation (continuous gray
line). The band energies from the effective model are plotted only
in the shaded regions, where approximation for the states is accurate
as discussed in Sec.~\ref{sec:effmodel}.} \label{fig:taylorbands3GK} 
\end{figure}

The accuracy of the Taylor expansion of the Hamiltonian matrices in
the previous subsection can be determined by comparing the band
energies obtained from our model with those from the \textit{ab
  initio} calculation.  In Fig.~\ref{fig:taylorbands3GK} we present
the band energies obtained from models including first-, second-, and
third-order expansions of the Hamiltonian on the lattice momentum
difference $\boldsymbol{\kappa}$; third-order expansions are not
discussed further in this work. We also show the \textit{ab initio}
bands for comparison, and focus on the regions where the approximation
for the states is accurate as discussed in
Sec.~\ref{sec:effmodel}. From Fig.~\ref{fig:taylorbands3GK}, we see
that keeping the cubic terms in the Hamiltonian expansion is
unnecessary to reproduce the \textit{ab initio} band energies around
the $\boldsymbol{\Gamma}$ point, while for the region around the
$\mathbf{K}$ point (and equivalently the $\mathbf{K}^{\prime}$ point)
it is actually detrimental to go beyond the second-order expansion.

\begin{figure}[htb!]
\centering
\begin{tabular}{rl}
 \includegraphics[width=0.49\columnwidth]{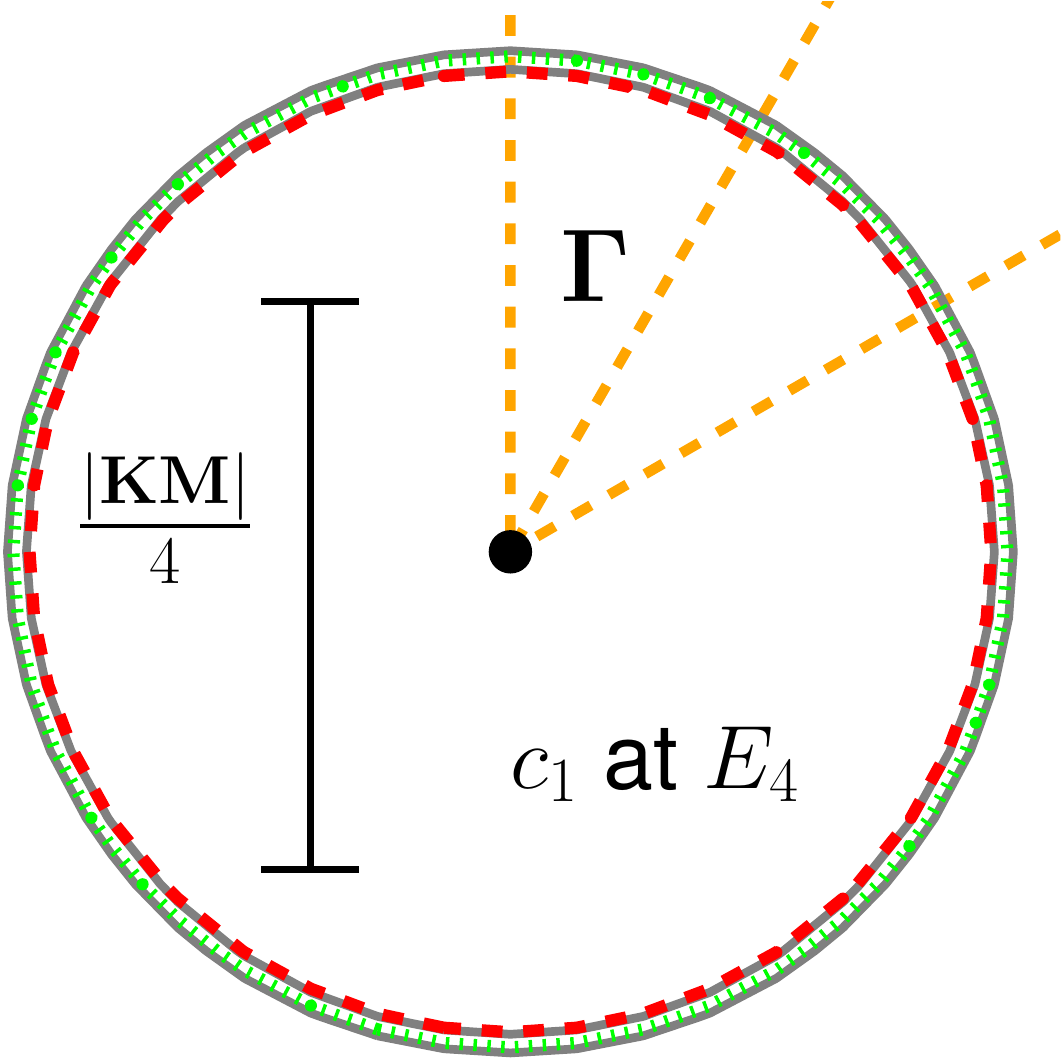} &
\includegraphics[width=0.49\columnwidth]{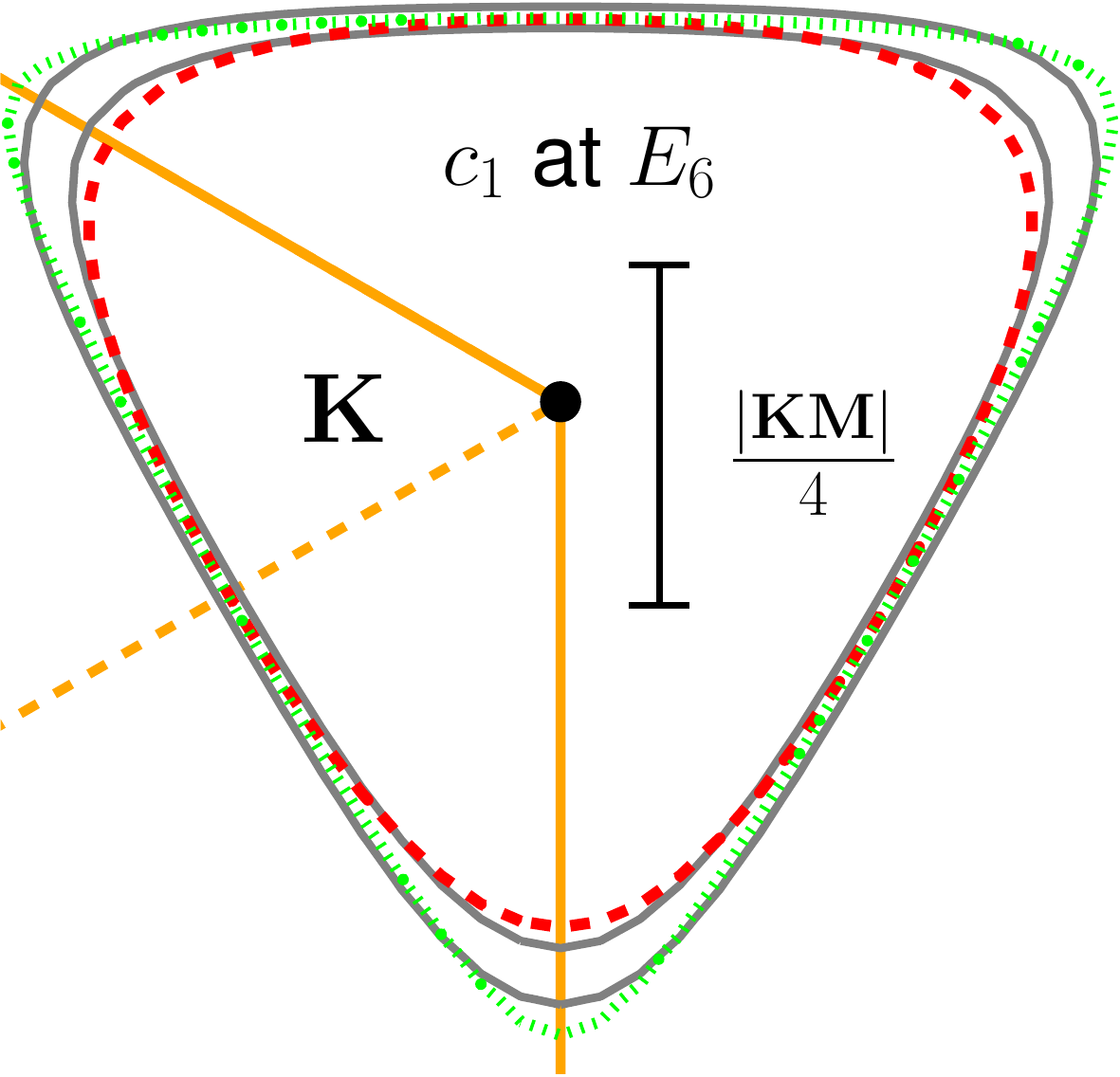} \\
%%\centerline{Fermi Level} \rule{0.85\columnwidth}{0.8pt}\\[1ex]
\includegraphics[width=0.49\columnwidth]{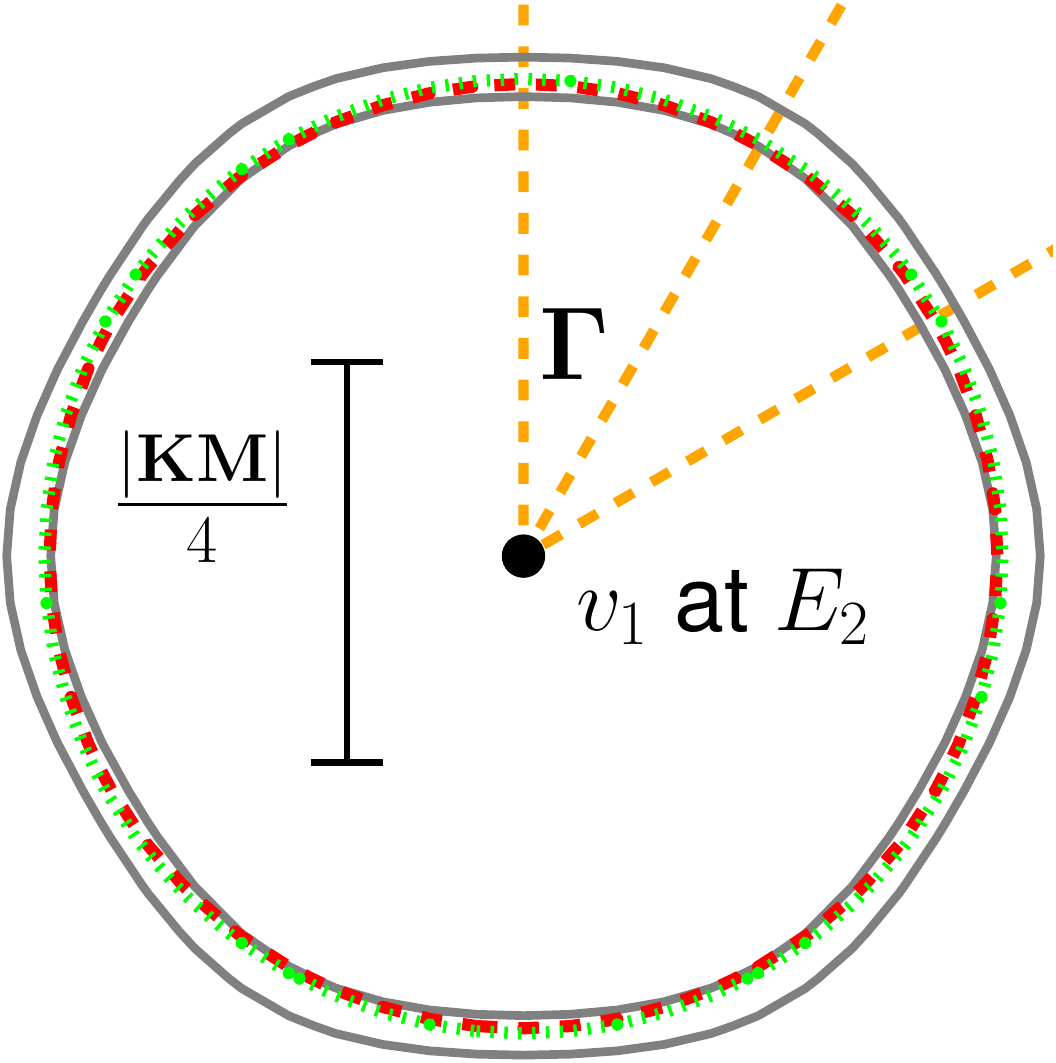} &
\includegraphics[width=0.49\columnwidth]{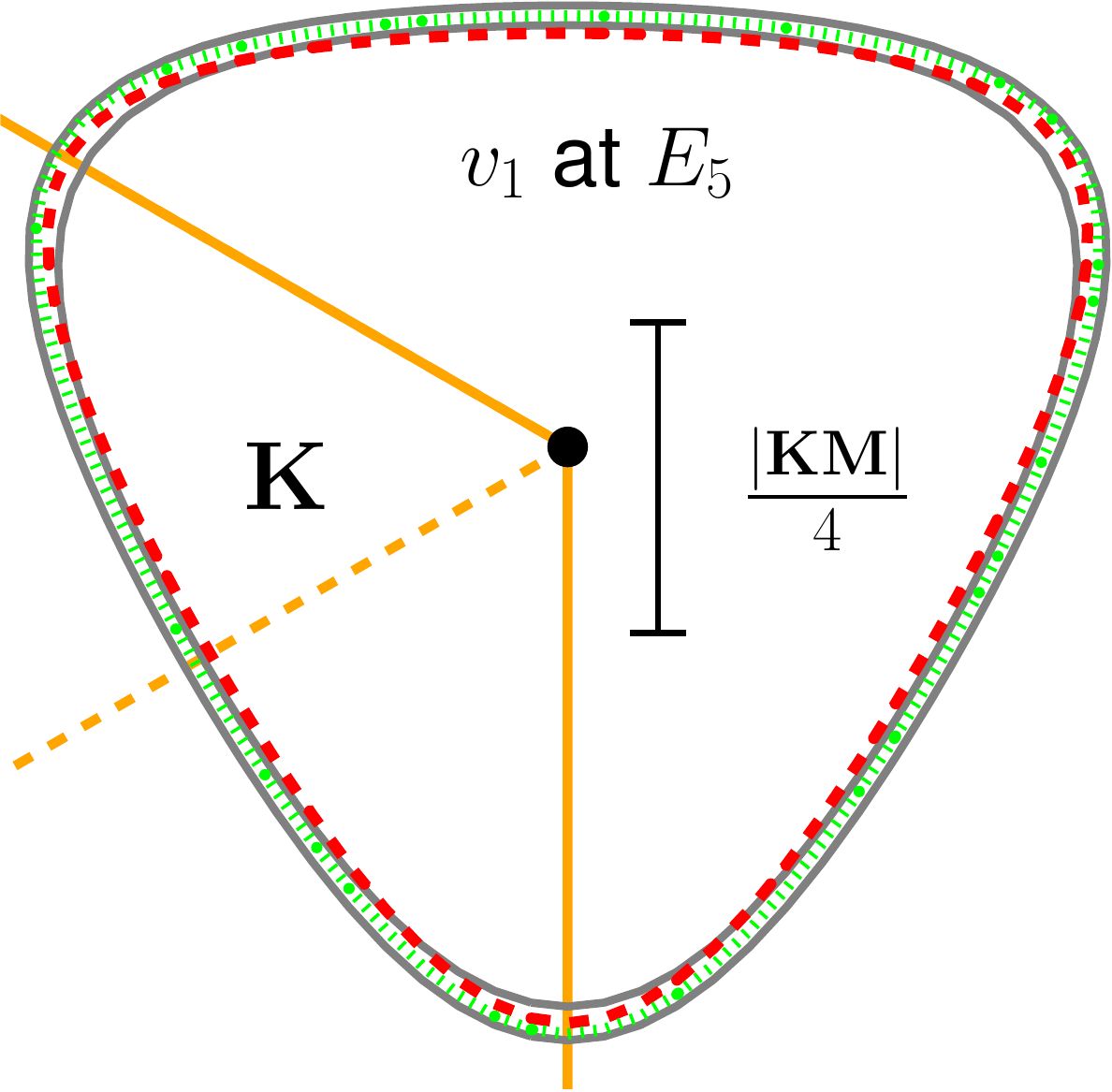} \\
 \includegraphics[width=0.49\columnwidth]{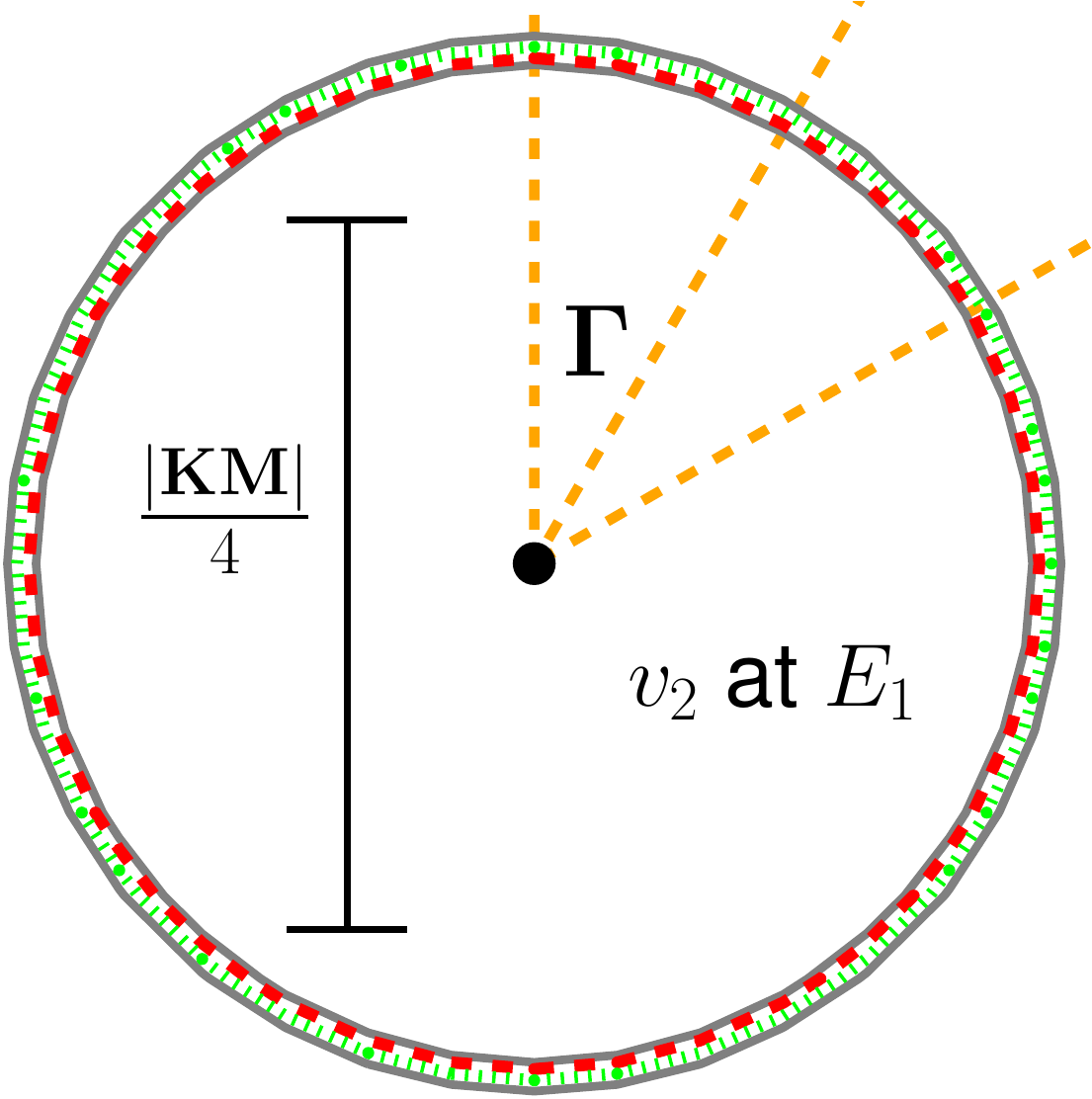} &
\end{tabular} 
\caption[]{ (Color online) Trigonal warping of the relevant bands
  around the reference points in the BZ in our model. Red-dashed and
  green-dotted lines indicate respectively second and third order
  expansions of the Hamiltonian matrices. The thin gray lines are
  \textit{ab initio} energy isolines that enclose a range of energy
  equivalent to room temperature. To give a sense of proportion we
  include a line segment of length one fourth of the distance
  $\overline{\mathbf{KM}}$. The isolevels $E_i$ and band labels $v_i$
  and $c_1$ are as indicated in Fig.~\ref{fig:taylorbands3GK}.}
\label{fig:trigwarp}
\end{figure}

A plot of band energies along a simple path through a region of the BZ
is not enough to establish the accuracy of the bands from our
model in that entire region. Analyzing the band warping is a way to
ensure that the good agreement displayed in
Fig.~\ref{fig:taylorbands3GK} is not coincidental to the directions
associated with that plot. In Fig.~\ref{fig:trigwarp} we show
isoenergy lines for each relevant band obtained from our model and
those from the \textit{ab initio} computation. The latter are shown as
pairs of lines that enclose an energy range equivalent to room
temperature, which is taken as our tolerance for energy accuracy.  We
compare the band warping corresponding to expansions of the
Hamiltonian that are quadratic and cubic on the lattice momentum
difference $\boldsymbol{\kappa}$; on Fig.~\ref{fig:trigwarp} we show
that the cubic expansion does not improve upon the quadratic one. Thus
we confirm that the quadratic expansion provides the best model for
the bandstructure of stanene for excitation energies up to $1.1\ e$V.

\section{Optical properties }
\label{sec:conducti}

\begin{figure}[htb!]
\centering \includegraphics[width=0.9\columnwidth]{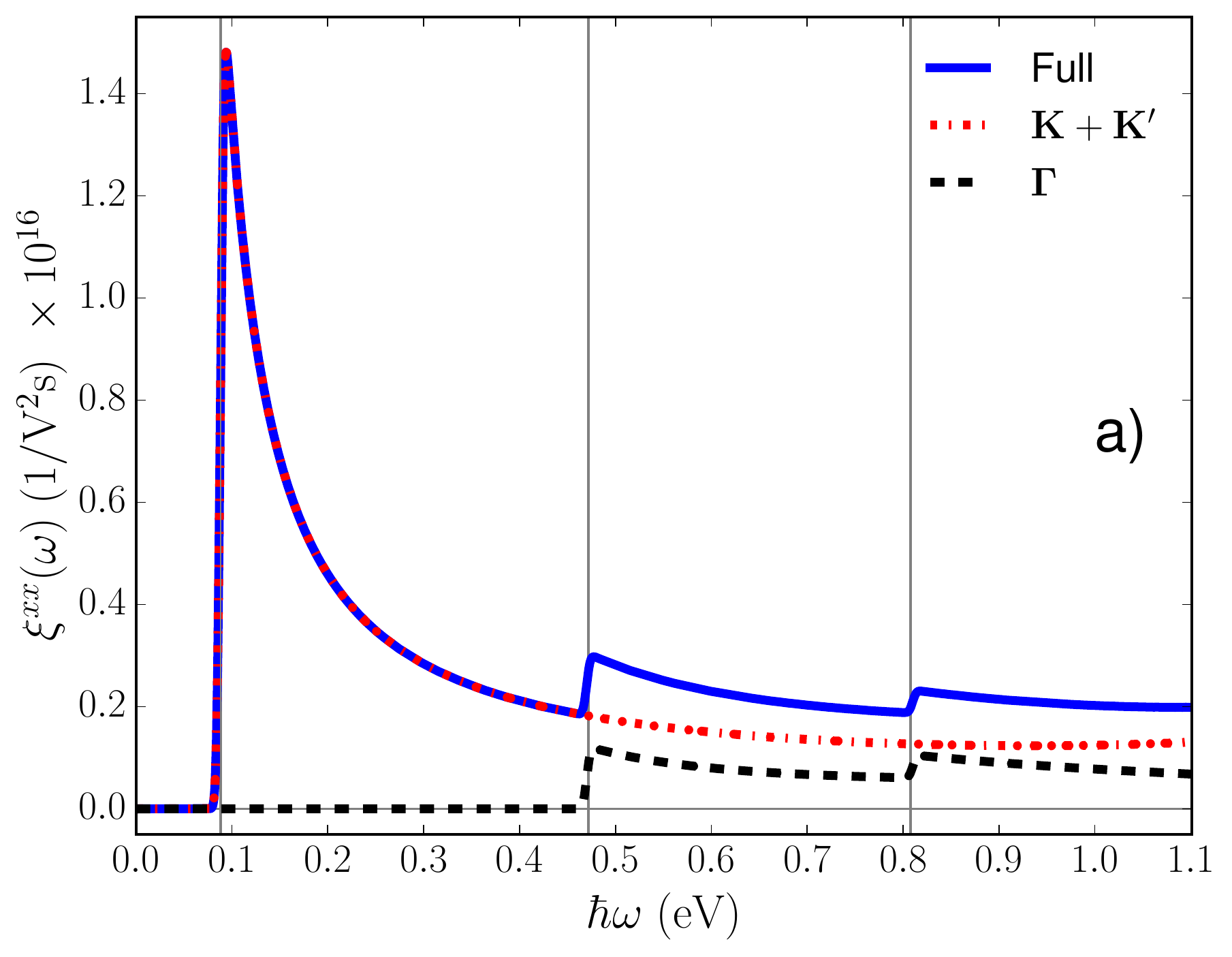}
\includegraphics[width=0.9\columnwidth]{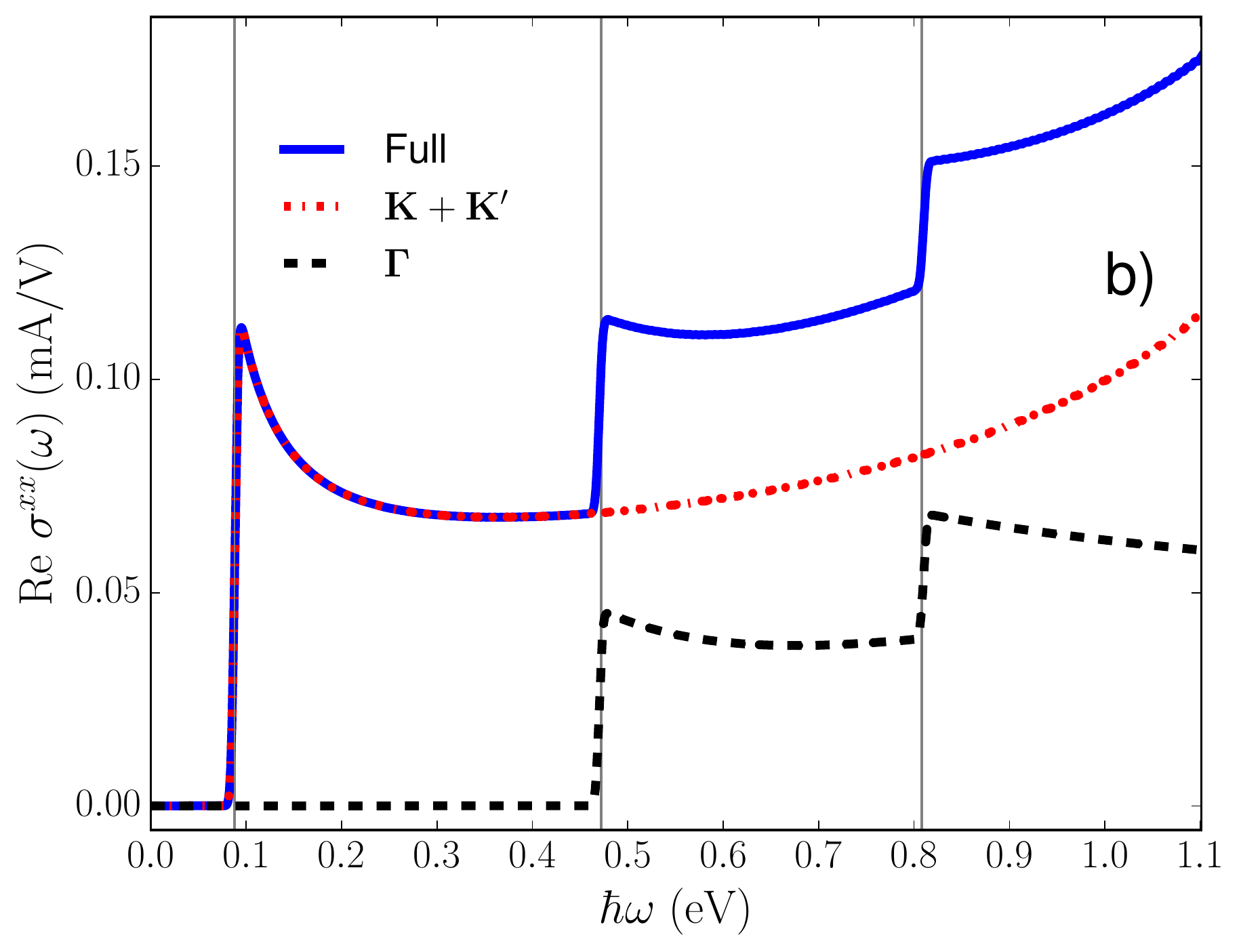} 
\caption[]{ (Color online) Linear optical absorption properties
  computed with the effective model of Sec.~\ref{sec:effmodel}. a)
  One-photon absorption coefficient $\xi^{xx}\left(\omega\right)$
  and b) real part of the optical conductivity
  $\sigma^{xx}\left(\omega\right)$ of stanene.  The contributions from
  the regions around the $\mathbf{K}$ and $\mathbf{K}^{\prime}$ points
  (dot-dashed red line) and the $\bm{\Gamma}$ point (dashed black line) in
  the BZ are shown separately, along with that from the full BZ (solid
  blue line);  we stress that the ``full'' signal indeed contains
  contributions from all crystal momenta $\bk$ around $\bK(\bKp)$ and
  $\bG$ for which one-photon transitions less than 1.1~eV are
  possible; consequently, it is equivalent to a full BZ calculation,
  within the limits of validity of our model. }
\label{fig:OPA} 
\end{figure}

The optical properties of a crystalline system depend only on the
Hamiltonian matrix and the Lax connection \cite{Muniz17}. Since the
Lax connection is null in the basis of our model
$\bar{\boldsymbol{\xi}}_{ab\bk}=0$, the velocity matrix elements are
simply given by
$\bm{v}\left(\mathbf{k}\right)=\hbar^{-1}\boldsymbol{\nabla}_{\mathbf{k}}\bar{H}\left(\mathbf{k}\right)$.
We consider the optical injection rates of carrier and spin densities,
given by
\begin{align}
\frac{d}{dt}n&= \xi^{ab}\left(\omega\right)E^{a}\left(\omega\right)\,E^{b}\left(-\omega\right),\label{eq:2}\\
\frac{d}{dt}S^z&= \zeta^{zab}\left(\omega\right)E^{a}\left(\omega\right)\,E^{b}\left(-\omega\right),
\end{align}
where we use the convention of summing  repeated indices, ${\bf E}\left(t\right)={\bf E}\left(\omega\right)e^{-i\omega t}+c.c.$
is an incident optical field, and the tensors $\xi^{ab}\left(\omega\right)$
and $\zeta^{zab}\left(\omega\right)$ are the carrier and spin
density injection coefficients 
\begin{align}
\xi^{ab}\left(\omega\right)&= \dfrac{2\pi e^{2}}{\hbar^{2}\omega^{2}}\underset{cv}{\sum}\int\dfrac{d^{2}k}{\left(2\pi\right)^{2}}v_{cv{\bf k}}^{a}v_{vc{\bf k}}^{b}\delta\left(\omega-\omega_{cv{\bf k}}\right),\label{eq:OPA}\\
\zeta^{zab}\left(\omega\right)&= \dfrac{2\pi e^{2}}{\hbar^{2}\omega^{2}}\underset{cv}{\sum}\int\dfrac{d^{2}k}{\left(2\pi\right)^{2}}\left(S_{cc}^{z}-S_{vv}^{z}\right)v_{cv{\bf k}}^{a}v_{vc{\bf k}}^{b}\delta\left(\omega-\omega_{cv{\bf k}}\right),\label{eq:SPA}
\end{align}
where $v$ and $c$ are respectively valence and conduction band indices,
$e=-|e|$ is the electron charge, $v_{cv{\bf k}}^{a}$ are the velocity
matrix elements, $S_{cc}^{z}=\pm\hbar/2$ and $S_{vv}^{z}=\mp\hbar/2$
are the spin-$\hat{{\bf z}}$ matrix elements of respectively the
conduction and valence bands, and $\hbar\omega_{cv\boldsymbol{k}}=\hbar\omega_{c\boldsymbol{k}}-\hbar\omega_{v\boldsymbol{k}}$
are band energy differences. In numerical calculations, we approximate
the Dirac delta function in the above equations by a Lorentzian function
with a broadening width of $6$~m$e$V.

In Fig.~\ref{fig:OPA}, we present plots of the linear optical
absorption coefficient $\xi^{xx}\left(\omega\right)$ and the real part
of the optical conductivity ${\rm Re}\sigma^{xx}\left(\omega\right)$,
which are related to each other by
$\xi^{xx}\left(\omega\right)=2\mathrm{Re\,}\sigma^{xx}\left(\omega\right)/\left(\hbar\omega\right)$.
As the frequency increases, the absorption begins at the band gap
energy $0.088\ e$V due to electronic transitions at the $\bK$ and
$\bK^{\prime}$ valleys in the BZ. The contribution from $\bG$ has an
absorption onset at $0.472\ e$V, and a second absorption onset at
$0.808\ e$V, when electronic transitions from the second valence band
are allowed.

\begin{figure}[htb!]
\centering \includegraphics[width=0.95\columnwidth]{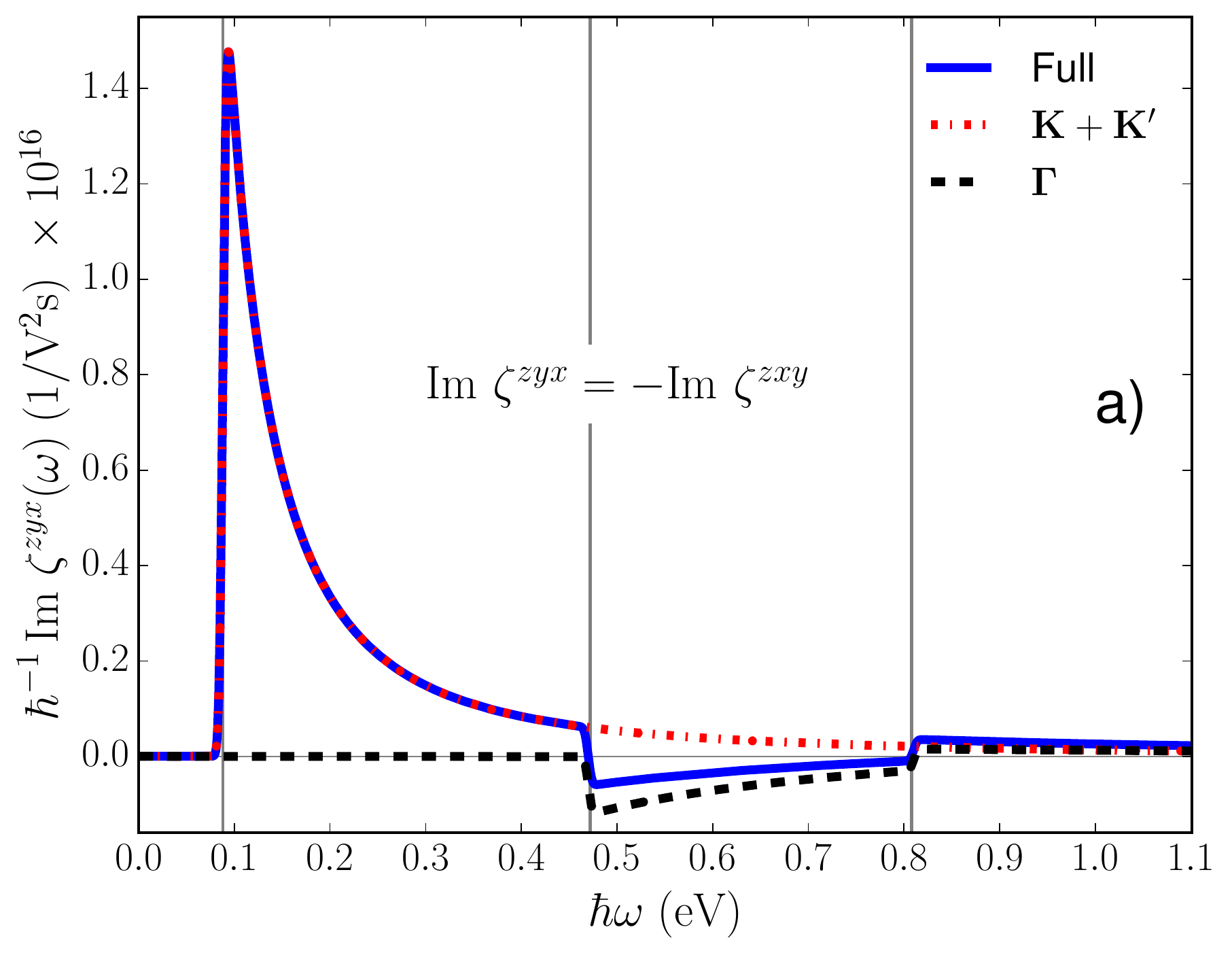}
\includegraphics[width=0.95\columnwidth]{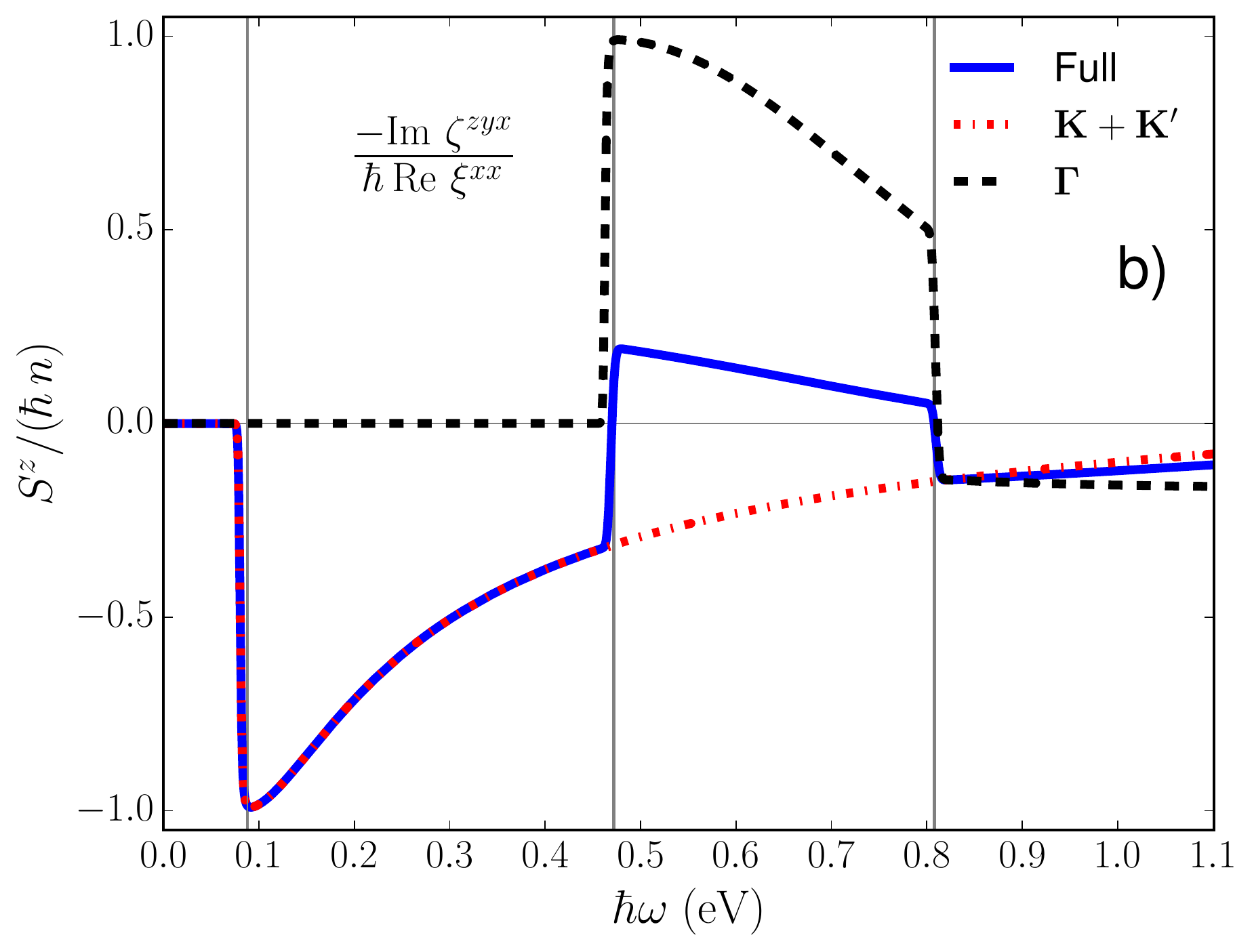} \caption[]{
  (Color online) Spin density injection computed with the effective
  model of Sec.~\ref{sec:effmodel}. a) Spin density injection
  coefficient $\zeta^{zyx}\left(\omega\right)$ and b) Spin
  polarization of injected carriers with circularly polarized light
  $-{\rm
    Im}\, \zeta^{zyx}\left(\omega\right)/\left[\hbar\xi^{xx}\left(\omega\right)\right]$
  for stanene, with $\mathfrak{h}=1$. The contributions from the
  regions around the $\mathbf{K}$ and $\mathbf{K}^{\prime}$ points
  (dot-dashed red line) and the $\bm{\Gamma}$ point (dashed black
  line) in the BZ are shown separately, along with that from the full
  BZ (solid blue line), in the sense described in the caption of
  Fig.~\ref{fig:OPA}.} \label{fig:spinPA}
\end{figure}

For photon energies close to the band gap, stanene has the interesting
property that circularly polarized light excites mostly electrons with the
spin that matches its helicity. Similar characteristics have
  been identified and studied in other monolayers, such as 
  silicene\cite{Ezawa_CircularDich}. This feature can be seen from our linear model
for the $\bK$ and $\bK^{\prime}$ points in Eq.~\ref{eq:H1K}, which can
be separated in spin sectors, and the expressions of
$\xi\left(\omega\right)$ and $\zeta\left(\omega\right)$ for a
Dirac cone \cite{muniz14,muniz15}.  For circular polarizations, the
light field propagating along the $\hat{{\bf z}}$ direction can be
written as
${\bf E}\left(\omega\right)=E_{\omega}\hat{{\bf p}}_{\mathfrak{h}}$,
where $\mathfrak{h}=\pm1$ is the helicity, and
$\hat{{\bf p}}_{\mathfrak{h}}=\left(\hat{{\bf
      x}}+i\mathfrak{h}\hat{{\bf y}}\right)/\sqrt{2}$.  Then the
expression for each spin $s=\pm1$ is
\begin{align}
\label{eq:step}
\xi_{\tau s}^{\mathfrak{h}}\left(\omega\right)= &
                                              \frac{\Theta\left(\omega-2\Delta_{K}\right)e^{2}}{8\hbar^{2}\omega}\left(1+\mathfrak{h} s\frac{2\Delta_{K}}{\omega}\right)^{2},
\end{align}
which is independent of valley, and where $\Theta\left(x\right)$ is
the step function, valued as zero or unity if $x<0$ or $x>0$,
respectively. From Eq.~\eqref{eq:step} we see that the spin
polarization is maximal for photon energies corresponding to the gap,
and it decreases for larger photon energies. The injection coefficient
of an arbitrary quantity for circularly polarized light,
$\eta^{\mathfrak{h}}\left(\omega\right)$, is given in terms of its
Cartesian components as
\begin{align}
\eta^{\mathfrak{h}}\left(\omega\right) &= \frac{1}{2}\left[\eta^{xx}\left(\omega\right)+\eta^{yy}\left(\omega\right)\right]+i\mathfrak{h}\frac{1}{2}\left[\eta^{yx}\left(\omega\right)-\eta^{xy}\left(\omega\right)\right]\notag\\
&= \eta^{xx}\left(\omega\right)+i\,\mathfrak{h}\,\eta^{yx}\left(\omega\right),
\end{align}
 where the relations $\eta^{xx}\left(\omega\right)=\eta^{yy}\left(\omega\right)$
and $\eta^{yx}\left(\omega\right)=-\eta^{xy}\left(\omega\right)$
due to the symmetries of a buckled honeycomb lattice were used. For carrier and spin densities
in stanene, we also have $\xi^{xx}\left(\omega\right)={\rm Re}\,\xi^{xx}\left(\omega\right)$
and $\xi^{yx}\left(\omega\right)=0$, as well as $\zeta^{zxx}\left(\omega\right)=0$
and $\zeta^{zyx}\left(\omega\right)=i\,{\rm Im}\,\zeta^{zyx}\left(\omega\right)$.
So the coefficients for circular polarizations are simply $\xi^{\mathfrak{h}}\left(\omega\right)={\rm Re}\,\xi^{xx}\left(\omega\right)$
and $\zeta^{\mathfrak{h}}\left(\omega\right)=-\mathfrak{h}\,{\rm Im}\,\zeta^{zyx}\left(\omega\right)$.
We present plots of the spin density injection coefficient $\zeta^{zxy}\left(\omega\right)$
computed with our effective model in Fig.~\ref{fig:spinPA} a), which
shows the same frequency regimes discussed for $\xi^{xx}\left(\omega\right)$.
In Fig.~\ref{fig:spinPA} b) we show the spin polarization of injected
carriers for circularly polarized light. Even for excitations at the
$\bG$ valley there is still a helicity-spin coupling, although the
net spin polarization is partially canceled by the excitations at
the $\bK$ and $\bK^{\prime}$ valleys. 

We note that helicity-spin coupling is due to the sign of the mass term
$\Delta_{K}$ in each Dirac cone \cite{ryu09,santos10,chamon12},
which  also explains why stanene shows the spin Hall effect. We
also point out that the helicity-spin coupling in stanene is analogous
to the helicity-valley coupling in TMDs \cite{Xiao-TMdchal}.

\section{Conclusions }
\label{sec:conclusion}

We have presented an effective model that accurately describes the
electronic and optical properties of stanene for low photon energies.
We started from an \textit{ab initio} calculation of the bandstructure
of stanene, which allowed us to identify the parameters in the model.
Our model includes a minimum set of energy states: 6 bands around
the $\bG$ point in the BZ, and 4 bands around the $\mathbf{K}$ and
$\mathbf{K}^{\prime}$ points. We provided measures for the accuracy
of the approximations for states and for energies, so we can identify
the range of validity of the model. 

We found that a quadratic model with respect to the lattice momentum
is the best suited for calculations based on the bandstructure. Even
the band warping from DFT calculations is better reproduced by the
quadratic rather than a cubic model. We also found that the lattice
buckling can be neglected.  This is confirmed by verifying that a
separation of the states according to spin-$\hat{{\bf z}}$ subsectors
is a good approximation for the band energies. In the Appendix, we
discuss the physical significance of some parameters in our model by
comparing it to a $p_{z}$-orbital tight-binding model expanded around
the ${\bf K}$ and ${\bf K}^{\prime}$ regions of the BZ. Finally, we
illustrated the applicability of the model by computing linear optical
absorption rates of stanene. We highlighted the coupling of circularly
polarized light with the electronic spin, which underscores the
potential of stanene for optical-spintronic applications.

The model proposed here can accurately describe optical properties of
stanene up to photon energies of $1.1\ e$V, which is suitable for a
wide range of optical experiments. Compared with a usual $\kdotp$
method, our model requires fewer parameters to describe the
bandstructure; we also provide a figure of merit to determine the
portion of the Brillouin Zone where the approximation is sensible. We
expect that this simple model will be useful in understanding and
suggesting experiments on this promising material, and that the
procedure described here will be used to extract effective models from
\textit{ab initio} calculations for other 2D materials.

\begin{acknowledgments}
  We thank Shu-Ting Pi (UC Irvine) for sharing a ONCVP pseudopotential
  for Sn atoms.  We acknowledge support from the Natural Sciences and
  Engineering Research Council of Canada (NSERC).
\end{acknowledgments}

\appendix

\section{Tight binding model }
\label{sec:TBmodel}

Tight-binding models (TBM) have successfully been used to describe
electronic states in the full Brillouin zone (BZ) in different
monolayer materials, such as silicene, germanene and
stanene\cite{Liu2011_2DHam,ezawa_silicene_efield}; the description of
the full BZ usually requires the inclusion of basis sets with $s$,
$p_{x}$, $p_{y}$ and $p_{z}$ orbitals. In this Appendix we discuss a
TBM that includes only $p_{z}$ orbitals on a buckled honeycomb
lattice, which is enough to describe the states around the
$\mathbf{K}$ and $\mathbf{K}^{\prime}$ points in the BZ, and in that
region it agrees with the models of Liu~\cite{Liu2011_2DHam} and
Ezawa~\cite{ezawa_silicene_efield}.  This TBM is unable to describe
the states at the $\boldsymbol{\Gamma}$ point because their orbital
character are not purely $p_{z}$. For instance, from a DFT
calculation, we find that at the $\boldsymbol{\Gamma}$ point the
orbital character of the first conduction band is 73\%~$s$,
24\%~$p_{z}$ and 3\%~$d$, while that of the top valence band is 96\% a
mix of $p_{x}$ and $p_{y}$, and about 4\% $d$ character. 
We assume that the orbitals are well localized and we apply a change
of basis to the $B$ sublattice
$u_{B\mathbf{k}}\left(\mathbf{r}\right)\to
e^{-i\mathbf{k}\cdot\boldsymbol{\delta}_{3}}u_{B\mathbf{k}}\left(\mathbf{r}\right)$
in order to have a basis with vanishing Lax connection.

This basis is not in Bloch's form\footnote{ When the periodic
  functions of a basis satisfy the condition
  $
  u_{\ell\bk+\mathbf{G}}\left(\mathbf{r}\right)=e^{-i\mathbf{G}\cdot\mathbf{r}}u_{\ell\bk}\left(\mathbf{r}\right),
  $ where $\mathbf{G}$ is a reciprocal lattice vector and $\ell$ is a
  band index, the Bloch
  wavefunctions are periodic over the Brillouin zone, 
  $
  \phi_{\ell\bk+\mathbf{G}}\left(\mathbf{r}\right)=\phi_{\ell\bk}\left(\mathbf{r}\right),
  $ and the basis is said to be in Bloch's form.}, and it allows us to
write all the hopping parameters in terms of the nearest neighbor
vectors $\boldsymbol{\delta}_{n}$ instead of the lattice vectors
$\boldsymbol{a}_{n}$. Using the notation employed in the main text,
the Hamiltonian is written in terms of the matrices
$s_{i}\otimes\sigma_{j}$.  With these conventions and employing a
usual tight-binding framework\cite{CastroNeto_RMP}, the
nearest-neighbour (NN) hopping term in the Hamiltonian is
\begin{equation}
\begin{array}{rl}
{\cal H}_{\mathbf{k}}^{\mathrm{NN}}= & -t\underset{n=1}{\overset{3}{\sum}}s_{0}\otimes\left[\begin{array}{cc}
0 & e^{-i\mathbf{k}\cdot\boldsymbol{\delta}_{n}}\\
e^{i\mathbf{k}\cdot\boldsymbol{\delta}_{n}} & 0
\end{array}\right]\\
= & -t\underset{n=1}{\overset{3}{\sum}}s_{0}\otimes\left[\cos\left(\mathbf{k}\cdot\boldsymbol{\delta}_{n}\right)\sigma_{x}+\sin\left(\mathbf{k}\cdot\boldsymbol{\delta}_{n}\right)\sigma_{y}\right],
\end{array}
\end{equation}
and the next-nearest-neighbor (NNN) term is 
\begin{equation}
\begin{array}{rl}
{\cal H}_{\mathbf{k}}^{\mathrm{NNN}}= & -t^{\prime}\underset{n, m\neq n}{\sum}\left[e^{i\mathbf{k}\cdot\left(\boldsymbol{\delta}_{m}-\boldsymbol{\delta}_{n}\right)}+e^{-i\mathbf{k}\cdot\left(\boldsymbol{\delta}_{m}-\boldsymbol{\delta}_{n}\right)}\right]s_{0}\otimes\sigma_{0}\\
= & -2t^{\prime}\underset{n, m\neq n}{\sum}\cos\left(\mathbf{k}\cdot\left(\boldsymbol{\delta}_{m}-\boldsymbol{\delta}_{n}\right)\right)s_{0}\otimes\sigma_{0},
\end{array}
\end{equation}
without the spin-orbit coupling. The spin-orbit coupling changes the
next-nearest-neighbor hopping matrices according to 
\begin{equation}
t^{\prime}s_{0}\otimes\sigma_{0}\to t^{\prime}s_{0}\otimes\sigma_{0}+i\lambda_{SO}\left(\hat{\boldsymbol{\delta}}_{m}\times\hat{\boldsymbol{\delta}}_{n}\right)\cdot{\bf s}\otimes\sigma_{z},
\end{equation}
 where $\lambda_{SO}$ is the spin-orbit coupling parameter. The last
term in the above equation can be further separated in two parts by
decomposing the $\hat{\boldsymbol{\delta}}_{m}\times\hat{\boldsymbol{\delta}}_{n}$
vector as 
\begin{equation}
\lambda_{SO}\hat{\boldsymbol{\delta}}_{m}\times\hat{\boldsymbol{\delta}}_{n}=\lambda_{z}\hat{{\bf z}}+\lambda_{b}\hat{{\bf z}}\times\left(\boldsymbol{\delta}_{m}-\boldsymbol{\delta}_{n}\right),
\end{equation}
where 
\begin{align}
\label{eq:lambdas}
\lambda_{z}&=\dfrac{a}{\sqrt{a^{2}+4b^{2}}}\lambda_{SO},\negthickspace & \lambda_{b}&=\dfrac{2b}{\sqrt{a^{2}+4b^{2}}}\lambda_{SO},  
\end{align}
according to the lattice buckling; the lattice parameters $a$ and
$b$ are depicted in Fig.~\ref{fig:lattices}.

In order to compare the tight-binding model with the one described in
Sec.~\ref{sec:TBmodel}, we now perform an expansion in powers of
$\boldsymbol{\kappa}$ around the $\bK$ and $\bK^{\prime}$ points in
the BZ, to which we respectively associate $\tau=+1$ and $\tau=-1$.
Here we do not consider the effect of a substrate, hence $\mu=0$ and
$\lambda_{R}=0$, so we are describing suspended stanene. Applying a
further change of basis to the $B$ sublattice,
$u_{B\mathbf{k}}\left(\mathbf{r}\right)\to ie^{i{\bf
    K}\cdot\boldsymbol{\delta}_{3}}u_{B\mathbf{k}}\left(\mathbf{r}\right)$,
the linear term is
\begin{equation}
\begin{array}{rl}
{\cal H}_{\tau\boldsymbol{\kappa}}^{\left(1\right)}= & -\tau\frac{9}{2}\lambda_{z}a^{2}s_{z}\otimes\sigma_{z}+\frac{3}{2}tas_{0}\otimes\left(\kappa_{x}\sigma_{x}+\tau\kappa_{y}\sigma_{y}\right)\\
 & +3t^{\prime}s_{0}\otimes\sigma_{0}-\frac{9}{2}\lambda_{b}a^{3}\left(\kappa_{y}s_{x}-\kappa_{x}s_{y}\right)\otimes\sigma_{z},
\end{array}\label{eq:TB1K}
\end{equation}
where the term $3t^{\prime}s_{0}\otimes\sigma_{0}$ is simply an energy
shift and can be removed. The quadratic term is 
\begin{equation}
\begin{array}{rl}
{\cal H}_{\tau\boldsymbol{\kappa}}^{\left(2\right)}= & -\frac{3}{4}ta^{2}s_{0}\otimes\left[\tau\kappa_{x}\kappa_{y}\sigma_{x}+\frac{1}{2}\left(\kappa_{x}^{2}-\kappa_{y}^{2}\right)\sigma_{y}\right]\\
 & -\frac{9}{4}t^{\prime}a^{2}\kappa^{2}s_{0}\otimes\sigma_{0}+\left(\frac{3}{2}\right)^{3}\lambda_{z}a^{4}\tau\kappa^{2}s_{z}\otimes\sigma_{z}\\
 & +\left(\frac{3}{2}\right)^{3}\lambda_{b}a^{4}\tau\left[\left(\kappa_{x}^{2}-\kappa_{y}^{2}\right)s_{x}-2\kappa_{x}\kappa_{y}s_{y}\right]\otimes\sigma_{z}.
\end{array}\label{eq:TB2K}
\end{equation}
 Now we compare this tight-binding model described by Eqs.~\eqref{eq:TB1K}-\eqref{eq:TB2K}
to our effective model around the $\bK$ and $\bK^{\prime}$ points
in the BZ described by Eqs.~\eqref{eq:H1K}-\eqref{eq:H2K}. The
relations between the respective first order parameters are 
\begin{align}
\Delta_{K}= & \tfrac{9}{2}\lambda_{z}a^{2}, & \zeta_{K}^{\left(1\right)}= & \tfrac{3}{2}t, & \lambda_{K}^{\left(1\right)}= & \tfrac{9}{2}\lambda_{b}a^{2},\label{eq:1}
\end{align}
 and for the second order ones, we have 
\begin{align}
\zeta_{K}^{\left(2\right)}= & \tfrac{3}{4}t, & v_{K}^{\left(2\right)}= & \tfrac{9}{4}t^{\prime}, & \vartheta_{K}^{\left(2\right)}= & \left(\tfrac{3}{2}\right)^{3}\lambda_{z}a^{2}, & \eta_{K}^{\left(2\right)}= & \left(\tfrac{3}{2}\right)^{3}\lambda_{b}a^{2},\label{eq:3}
\end{align}
 Since $\lambda_{b}=2b \lambda_{z} / a$, we can take $t$, $t^{\prime}$
and $\lambda_{z}$ to be the only independent parameters of the tight-binding
model;  numerical values for them can be obtained from Table I. Consequently, 
\begin{align}
\lambda_{K}^{\left(1\right)}= & \frac{2b}{a}\Delta_{K}, & \zeta_{K}^{\left(2\right)}= & \frac{1}{2}\zeta_{K}^{\left(1\right)}, & \vartheta_{K}^{\left(2\right)}= & \frac{3}{4}\Delta_{K}, & \eta_{K}^{\left(2\right)}= & \frac{3b}{2a}\Delta_{K}.\label{eq:4}
\end{align}
This tells us that $\Delta_{K}$, $\zeta_{K}^{\left(1\right)}$ and
$v_{K}^{\left(2\right)}$ can be taken as the only independent parameters
in Table~\ref{tab:paramK}, just as the 3 independent parameters
for the tight-binding. The parameters $t^{\prime}$ and $v_{K}^{\left(2\right)}$
can be neglected, though, so the relevant parameters are only two:
$t$ and $\lambda_{z}$ for tight-binding, and $\Delta_{K}$ and $\zeta_{K}^{\left(1\right)}$
in our effective model. The relations above are satisfied by the parameters
shown in Table~\ref{tab:paramK}.

\bibliography{references}

\end{document}